\documentclass[10pt,letterpaper]{article}

\usepackage{multirow}
\usepackage{multicol}
\usepackage{longtable}
\usepackage{amssymb}
\usepackage{paralist}
\usepackage{times}
\usepackage{pbox}
\usepackage{setspace}
\usepackage{graphicx}
\usepackage{algorithm}
\usepackage{algorithmic}
\usepackage{units}
\usepackage{microtype}
\usepackage{appendix}
\usepackage{hyperref}
\usepackage{color}
\usepackage{units}
\usepackage{microtype}
\usepackage{relsize}
\usepackage{verbatim}
\usepackage{soul}
\definecolor{MyBlue}{rgb}{0.2,0.2,0.8}

\newcommand{\ess}{\mathcal{S}}
\renewcommand{\H}{\mathcal{H}}

\newcommand{\I}{\mathcal{I}} 

\newcommand{\G}{\mathcal{G}}

\usepackage{graphicx}
\newcounter{lastnote}

\usepackage{booktabs}

\usepackage[top=0.85in,left=2.75in,footskip=0.75in]{geometry}

\usepackage{changepage}

\usepackage[utf8]{inputenc}

\usepackage{textcomp,marvosym}

\usepackage{fixltx2e}

\usepackage{amsmath,amssymb}

\usepackage{cite}

\usepackage{nameref,hyperref}

\usepackage[right]{lineno}

\usepackage{microtype}
\DisableLigatures[f]{encoding = *, family = * }

\usepackage{rotating}

\usepackage{tabularx}

\raggedright
\usepackage[aboveskip=1pt,labelfont=bf,labelsep=period,justification=raggedright,singlelinecheck=off]{caption}
\usepackage{subcaption}
\usepackage{multirow}
\usepackage{longtable}
\usepackage{multicol}
\usepackage{amssymb}
\usepackage{scicite}
\usepackage{paralist}
\usepackage{amsmath}
\usepackage{times}
\usepackage{pbox}
\usepackage{setspace}
\usepackage{caption}
\usepackage{algorithm}
\usepackage{algorithmic}
\usepackage{units}
\usepackage{microtype}
\usepackage{appendix}
\usepackage{hyperref}
\usepackage{color}
\usepackage{units}
\usepackage{microtype}
\usepackage{relsize}
\usepackage{verbatim}
\usepackage{multirow}
\usepackage{multicol}
\usepackage{amssymb}
\usepackage{scicite}
\usepackage{paralist}
\usepackage{times}
\usepackage{pbox}
\usepackage{setspace}
\usepackage{graphicx}
\usepackage{caption}
\usepackage{subcaption}
\usepackage{caption}
\usepackage{algorithm}
\usepackage{algorithmic}
\usepackage{units}
\usepackage{microtype}
\usepackage{appendix}
\usepackage{color}
\usepackage{units}
\usepackage{microtype}
\usepackage{relsize}
\usepackage{verbatim}
\usepackage{color}
\usepackage{soul}
\usepackage[show]{chato-notes} %%%%%%%%%% Chage to 'hide' before submission

\usepackage{graphicx}
\usepackage{booktabs}
\usepackage{setspace}

\makeatletter
\renewcommand{\@biblabel}[1]{\quad#1.}
\makeatother

\date{}

\usepackage{lastpage,fancyhdr,graphicx}
\fancyheadoffset[L]{2.25in}
\fancyfootoffset[L]{2.25in}
\begin{document}
\vspace*{0.35in}

% Please capitalize all terms in the title except conjunctions, prepositions, and articles.
\begin{flushleft}
{\Large
\textbf\newline{Prediction and Characterization of High-Activity Events
in Social Media Triggered by Real-World News}
}
\newline
\\
Janani Kalyanam\textsuperscript{1,*},
Mauricio Quezada\textsuperscript{2},
Barbara Poblete\textsuperscript{2},
Gert Lanckriet\textsuperscript{1}
\\
\bigskip
\bf{$^1$} Department of Electrical and Computer Engineering \\ University of California, San Diego, California, U.S.A.
\\
\bf{$^2$} Department of Computer Science \\ University of Chile, Santiago, Chile
\\
\bigskip

* jkalyana@ucsd.edu

\end{flushleft}
\section*{Abstract}

On-line social networks publish information on a high volume of
real-world events almost instantly, becoming a primary source for
breaking news.  Some of these real-world events can end up
having a very strong impact on on-line social networks.  The effect of such
events can be analyzed from several perspectives, one of them being
the intensity and characteristics of the collective activity that it
produces in the social platform.

We research 5,234 real-world news events encompassing 43 million
messages discussed on the Twitter microblogging service for
approximately 1 year.  We show empirically that exogenous news events
naturally create
collective patterns of bursty behavior in combination with long periods of
inactivity in the network. This type of behavior agrees with
other patterns previously observed in other types of natural collective phenomena, as
well as in individual human communications. In addition, we propose a methodology to
classify news events according to the different levels of intensity in
activity that they produce. In particular, we analyze the most
highly active events and 
observe a consistent and strikingly different collective reaction
from users when they are exposed to such events.  This reaction is
independent of an event's reach and scope.  We further observe that
extremely high-activity events have characteristics that are quite distinguishable
at the beginning stages of their outbreak.  This allows us to predict
with high precision, the top 8\% of events that will have the most
impact in the social network by just using the first 5\% of the information of an
event's lifetime evolution. This strongly implies that high-activity events
are naturally prioritized collectively by the social network, engaging users early
on, way before they are brought to the mainstream audience.

\section*{Introduction}
%\section{a}

% Motivation
Social media is now a primary source of breaking news information
for millions of users all over the world \cite{Kwak:2010}. On-line
social networks along with mobile internet devices have crowdsourced
the task of disseminating real-time information. As a result, both
news media and news consumers have become inundated with much more
information than they can process. One possible way of handling this data overload, is
to find ways to filter and prioritize information that has the
potential of creating a strong collective impact. Understanding and
quickly identifying the type of reaction that certain exogenous events will produce
in on-line social networks, at both global and local scales, can help
in the understanding of collective human behavior, as well as
improve information delivery, journalistic coverage and
crisis management, among other things. We
address this challenge by analyzing the properties of real-world news
events in on-line social networks, showing that they corroborate patterns
previously identified in other case studies of human communications. In
addition, we present our main findings of how news events that produce
extremely high-activity can be clearly identified in the early stages of
their outbreak.

% Brief background on the problem

The study of information propagation on the Web has sparked tremendous
interest in recent years. Current literature on the subject primarily
considers the process through which a {\em meme}, usually a piece of
media (like a video, an image, or a specific Web article), gains
popularity
\cite{Castillo:2014,Szabo:2010,Lerman:2010,Tatar2014,Pinto:2013,Ahmed:2013,Li:2016:concept:drift,
Liu:2015:UN}.  
However, a meme represents a
simple information unit and its propagation behavior does not necessarily
correspond to that of more complex information such as
news events. News events are usually diffused in the network in many
different formats, e.g., a particular news story such as an {\em
  earthquake in Japan} can be communicated through images, URLs,
tweets, videos, etc. Therefore, current research can benefit from analyzing
the effects of more high-level forms of information. 

Traditionally, the impact of information in on-line social networks has been
measured in relation to the total amount of attention that this subject receives
\cite{berger2012makes,iribarren2011branching,guerini2011exploring,mills2012virality,gaugaz2012predicting}.
That is, if a content posted in the network receives
votes/comments/shares above a certain threshold it is usually deemed as {\em viral} or
{\em popular}. Nevertheless, this
notion of popularity or impact will favor only information that produces very large
volumes of social media messages. 
Naturally, global breaking news that has world-wide coverage and that produces a high volume of
activity in a short time should be considered as
having a strong impact on the network.  However, there are other types of events
that can produce a similar reaction in smaller on-line communities
such as, for example, on users from a particular country
(e.g., the
withdrawal of the main right wing presidential candidate in Chile due
to psychiatric problems, just before
elections \cite{chile_elections}).
Clearly, events of local scope do not produce as much social media
activity as events of global scope, but they can create a strong and
immediate reaction from users in local networks \cite{ReisBOPKA15}. Conversely,
there are large events which do not produce an intense reaction, such as
{\em The Oscars} (Fig.~\ref{fig:fig1b}), which span a long
period of time and are discussed by social network users for weeks or
even months, but do not spark intense user activity. Therefore, it is reasonable to consider additional dimensions,
than just volume, when analyzing the impact of information in on-line communities.  

Prior research has shown that certain types of individual activities,
such as communications (studied in email exchanges), work patterns and
entertainment, follow a behavior of bursts of rapidly occurring
actions followed by long periods of inactivity
\cite{barabasi2005origin}, referred to as {temporally inhomogeneous}
behavior \cite{karsai2012universal}.  This type of behavior initially
observed in individual activities, has also been observed in relation
to other naturally occurring types of collective phenomena in human
dynamics similar to processes seen in self-organized criticality
\cite{karsai2012universal}.  In particular, extremely high-activity
bursty behavior seems to also occur in critical situations, observed
from the information flow in cell phone networks during
emergencies\cite{gao2014quantifying}.  Although, there is research
towards modeling this type of collective behavior
\cite{yan2013information} in on-line social networks, to the best of
our knowledge, it has not yet been analyzed quantitatively.

% when
% consulting journalists on how news media sources measure the impact of
% news, we learn that they too face the issue of not having a clear way
% to approach this problem.

% Our contributions

%\newtext{

Our work focuses on high-activity events in social media produced by
real-world news, with the following contributions:
\begin{enumerate}

\item We introduce a methodology for modeling and classifying
events in social media, based on the intensity of the activity that they
produce. This methodology is independent of the size and scope of the event,
and is an indicator of the impact that the event information had on the social network.

\item We show empirically that real-world news events produce collective
patterns of bursty behavior in the social network, in combination with long periods of
inactivity. Furthermore, we identify events for which most of their activity
is concentrated into very high-activity periods, we call these events {\em
high-activity events}.

\item We determine the existence of unique characteristics that
differentiate how high-activity events propagate in the social network.

\item We show that an important portion of high-activity events can be
predicted very early in their lifecycle, indicating that this type
of information is spontaneously identified and filtered collectively, early
on, by social network users.

\end{enumerate}
%}

%We focus on high-impact news events in social media, contributing by (i)
%\textcolor{blue}{defining a new concise way for measuring information impact that
%is independent of the size (whether large or small) and scope (whether
%local or global) of the event, but is representative of the urgency
%and immediacy of the reaction of users on the social media} (ii)
%determining the existence of unique characteristics that differentiate
%how high-impact exogenous events are propagated in the network, and
%(iii) show, through the creation of an early prediction model for
%high-impact events, that these types of news events are naturally
%identified and filtered by the network at very early stages of their
%outbreak.

\section*{Materials and Methods}

% model
We define an event as a conglomerate of information that encompasses
all of the social media content related to a real-world news
occurrence. Using this specification, which considers an event as a
complex unit of information, we study the type of collective reaction
produced by the event on the social network. In particular, we analyze 
the intensity or immediacy of the social network's response. 
By analyzing the levels of intensity in activity induced by different exogenous
events to the network, we are implicitly studying the priority that has been
collectively assigned to the event by groups of
independent individuals \cite{barabasi2005origin, karsai2012universal}. 

We characterize an event's discrete activity dynamics by using
\emph{interarrival times} between consecutive social media messages
within an event (e.g., $d_i = t_{i+1}-t_i$, where $d_i$ denotes the
interarrival time between two consecutive social media messages $i$
and $i+1$ that arrived in moments $t_i$ and $t_{i+1}$, respectively).

We introduce a novel vectorial representation based on a {\em vector
quantization of the interarrival time distribution}, which we call 
{\em ``VQ-event model''}. This model is
designed to filter events based on the distribution of the
interarrival times between consecutive messages.  This approach is inspired
by the {\em codebook-based representation} from the field of multimedia
content
analysis, which has been used in audio processing and computer vision
~\cite{ff,Vaizman}.  In our proposed approach, our method learns a set of
the most
representative interarrival times from a large training corpus of events;
each one of the representative interarrival times is known as a
{\em codeword} and the complete learned set is known as the {\em
codebook}~\cite{Vaizman}.  
Each event is then modeled using a vector quantization (VQ) that
converts the interarrival times of an event into a discrete set of values,
each value corresponding to the closest codeword in the codebook (details
in supplementary material).  The resulting VQ-event model is then a
vector in which each dimension contains the percentage of interarrival times
of the event that were assigned a particular codeword in the codebook.

The VQ-event representation is relative to an event's overall size
since the model is normalized with respect to the number of messages in the
event. Therefore the only criteria that are considered in the model are the
interarrival times of each particular event. This model allows us to group events based on the
{\em similarity of the distribution} of their interarrival times.
In those terms, we consider as high-activity events those events for which
the distribution of interarrival times is most heavily
skewed towards the smallest possible interval, zero.  In other words,
events for which the overall activity is extremely intense in comparison
with other events.

\begin{figure}
  \centering
  \begin{subfigure}{\textwidth}
    \includegraphics[width=\textwidth]{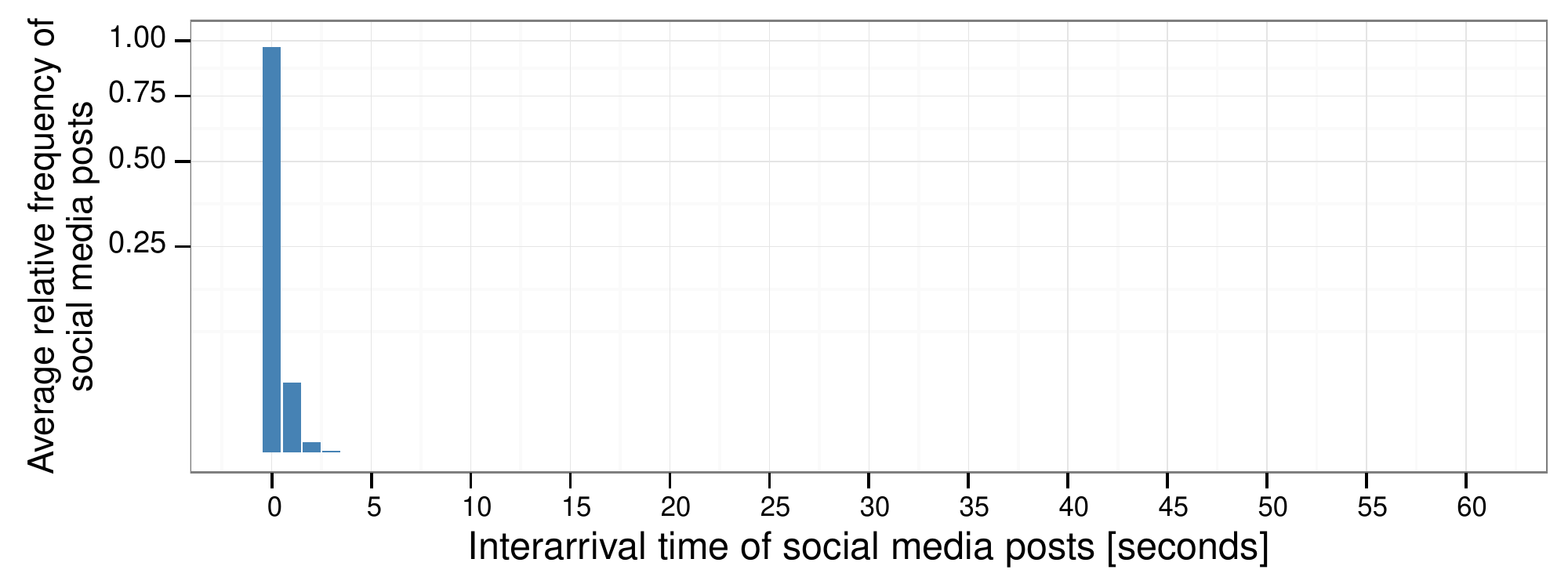}
    \caption{User posts about the death of Nelson Mandela arrive
      almost instantly.}
    \label{fig:fig1a}
  \end{subfigure}%

  ~ %add desired spacing between images, e. g. ~, \quad, \qquad, \hfill etc.
  % (or a blank line to force the subfigure onto a new line)
  \begin{subfigure}{\textwidth}
   \includegraphics[width=\textwidth]{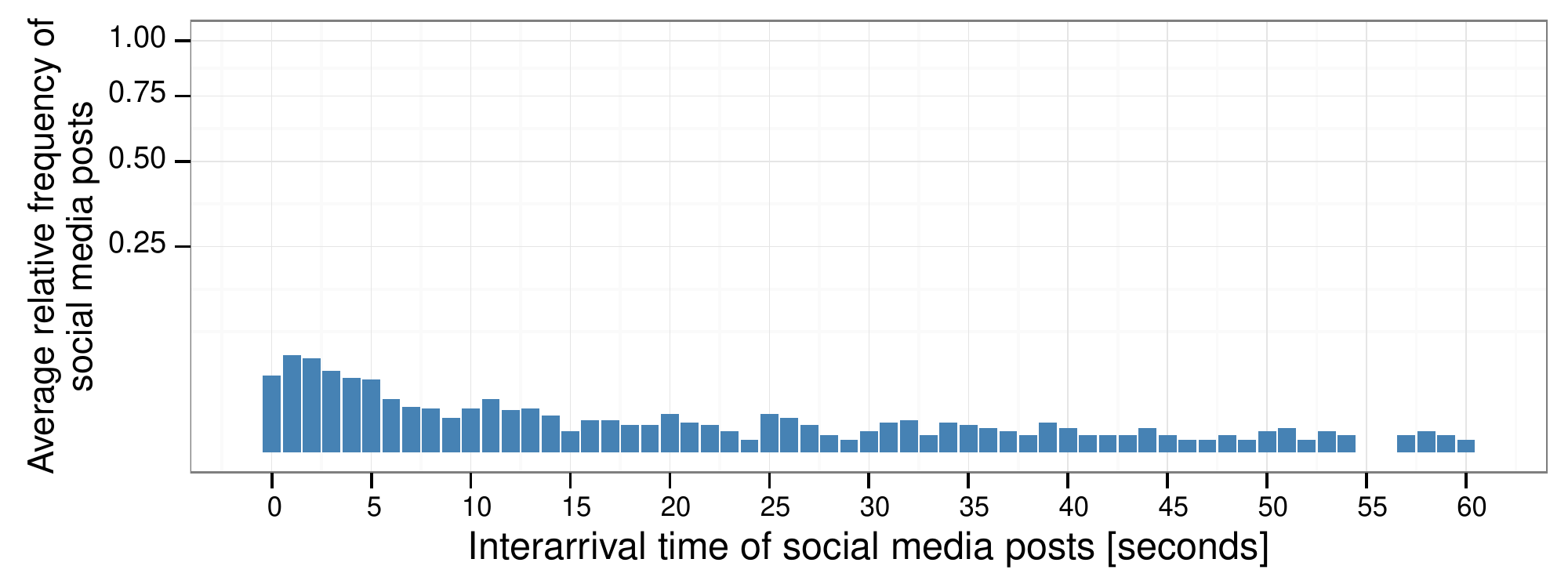}
    \caption{User posts about The Oscars arriving several weeks before
      the event.}
    \label{fig:fig1b}
  \end{subfigure}%
  ~ %add desired spacing between images, e. g. ~, \quad, \qquad, \hfill etc.
  % (or a blank line to force the subfigure onto a new line)

  \caption{\textbf{Examples of interarrival time histograms of two real-world news
events discussed on Twitter. The event [nelson, mandela] (\ref{fig:fig1a}) was
      collected on 12/05/2013. Since there is a high
      concentration in the first histogram bin, we conclude that most of the social media posts
      for this event occur in one or more successions of high-activity
      bursts (therefore, considered a high-activity event).
      The second event, [may, oscar] (\ref{fig:fig1b}) was collected
      on 03/23/2014 about The Oscars event that was held a few
      weeks before. The arrival times of these posts are much more spread
      out, displaying much less concentration of bursty activity.} 
  }
  \label{fig:fig1}
\end{figure}

To illustrate events with different levels of intensity in activity we
present two examples taken from our analysis of Twitter data. These
examples show the interarrival time histograms for the entire lifecycle of
the two events. In
the first example, the majority of the messages about
the death of political leader Nelson Mandela
(Fig.~\ref{fig:fig1a}) arrive within almost zero seconds of
each other. On the contrary, the messages about The Oscars
(Fig.~\ref{fig:fig1b}) are much more spread out in time.

We note that, by using interarrival times to describe the intensity of the
activity of an event, we make our analysis independent of the particular
evolution of each event. By doing this, we put no restrictions on how
high-activity events unfold in time, for example, they could be: (a) events
that start out slowly and
suddenly gain momentum, (b) events that go viral soon after they appear on
social media and then decay in intensity over a long (or short) period of
time, (c) events that from the beginning produce large amounts of interest and
sustain that interest throughout their long (or short) lifespan, or (d)
events that are a concatenation of any of the above, etc.

%\newtext{ 
% Experimental analysis

We study a dataset of news events gathered from news
headlines from a \emph{manually curated} list of well-known news media
accounts (e.g., @CNN, @BreakingNews, @BBCNews, etc.) in the
microblogging platform Twitter \cite{Twitter_website}
%\footnote{\url{https://twitter.com}
%  (Accessed: August 25, 2015.)} 
(a full list of all the news media
accounts is provided in the supplementary material). Headlines were
collected periodically every hour, over the course of approximately
one year. In parallel, all the Twitter messages (called \emph{tweets})
were extracted for each news event using the public
API \cite{Twitter_API}.
%\footnote{\url{https://dev.twitter.com/} (Accessed: August 25,
%  2015.)},
% In this research, since we focus on the microblogging platform
% Twitter, we collected all the Twitter messages (called
% \emph{tweets}) produced about each news event using the publicly
% available Twitter Search API
This process was performed by automatically extracting descriptive
sets of keywords for each event using a variation of frequent itemset
extraction \cite{Tan_Steinbach_Kumar} over the event's headlines.
These sets of keywords were then used to retrieve corresponding user
tweets for each event. We validate the events gathered in our
data collection process to ensure that each group of social media
posts corresponds to a meaningful and cohesive news event. We provide a detailed
description of the collection methodology and of the validation of event
cohesiveness in the supplementary material. Overall, the resulting dataset contains
$43,256,261$ tweets that account for $5,234$ events (Table~\ref{table:dataset-stats}).

\begin{figure}
    \includegraphics[width=\textwidth]{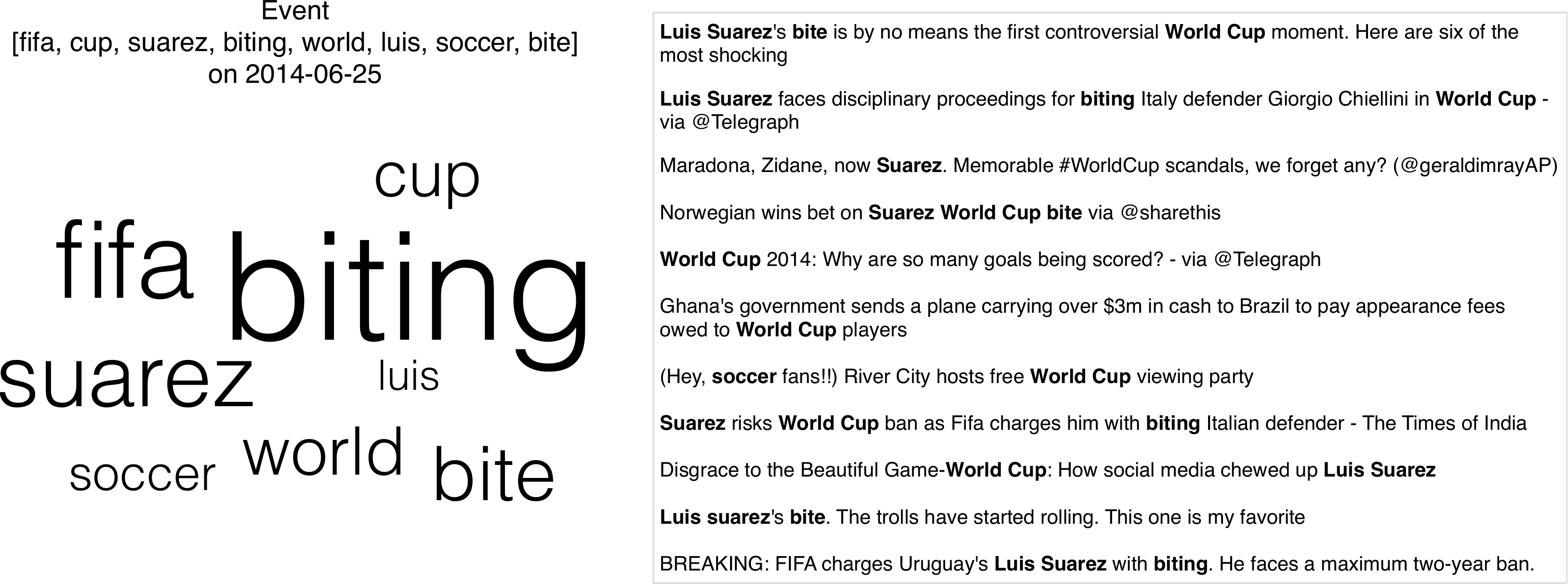}
  \caption{\textbf{An example event, collected on 06/25/2014
      with keywords (left) and sample user posts (right) obtained
      from the Twitter Search API. The tweets in the event contain at
      least a pair of descriptive keywords and were retrieved close to the time
      of the event.}}
  \label{fig:fig2}
\end{figure}

In Figure~\ref{fig:fig2} we characterize an example event from our
dataset, by showing the set of keywords and a sample of tweets
associated to the event. These keywords form a semantically meaningful
event; they refer to the incident where soccer player Luis Suarez was
charged for biting another player during the FIFA World Cup in
2014. This general collection process results in a set of social media
posts associated to an event which can encompass several memes, viral
tweets and pieces of information. Therefore, an event is composed of
diverse information, addressing more heterogeneous content than prior
work
% \cite{Castillo:2014},\cite{Szabo:2010},\cite{Lerman:2010},\cite{Tatar:2011},\cite{Pinto:2013},\cite{Ahmed:2013,
% Zaman_information_spreading},\cite{suh2010want}
\cite{Castillo:2014,Szabo:2010,Lerman:2010,Tatar:2011,Pinto:2013,Ahmed:2013,suh2010want}
which focus on single pieces of information (e.g., a
particular meme, a viral tweet etc.).
\begin{table}
  \centering
  \begin{tabularx}{\textwidth}{@{}p{6cm}llll@{}}
    \toprule
    \textbf{Event Collection Statistics} & \textbf{Minimum} & \textbf{Mean} & \textbf{Median} & \textbf{Maximum} \\ \midrule
    \# of posts (per event) & 1,000 & 8,254 & 2,474 & 510,920 \\
    \# of keywords (per tweet) & 2 & 3.77 & 3 & 39 \\
    Event duration (hours) & 0.12 & 20.93 & 7.46 & 190.43 \\ \bottomrule
  \end{tabularx}
  \caption{\bf High-level description of the dataset of news events.} \label{table:dataset-stats}
\end{table}

\begin{figure}
   \includegraphics[width=\textwidth]{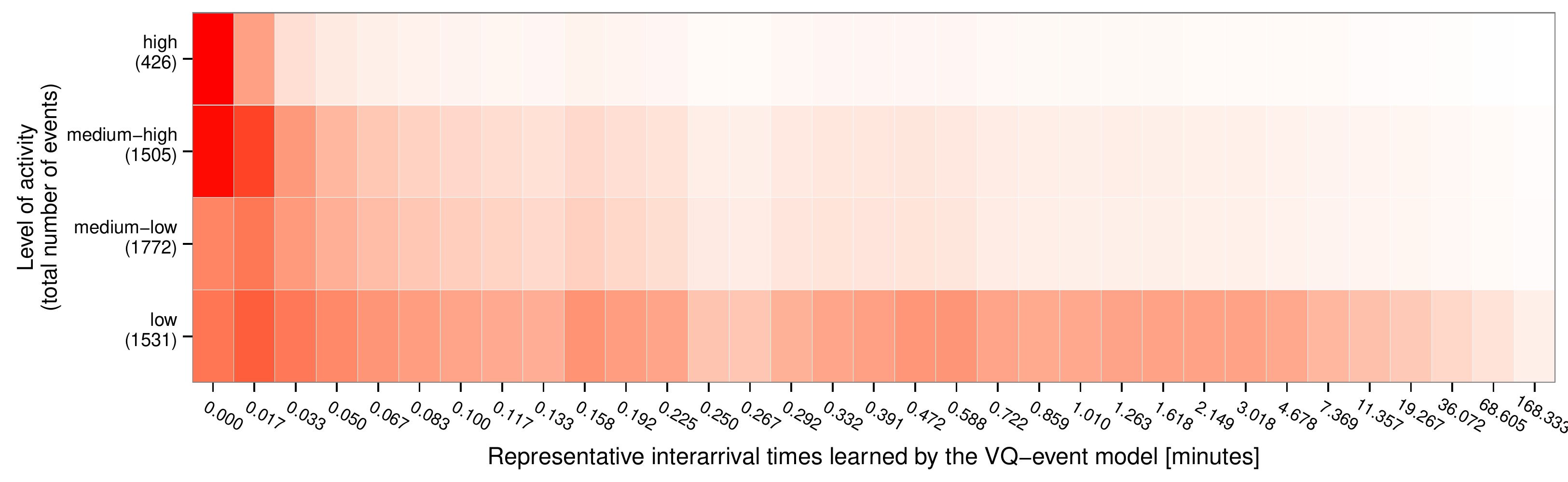}
  \caption{\textbf{Each row is the average representation of all the
      events in a cluster.  A darker cell represents a higher
      relative frequency value.  The y-axis specifies the number of events in
      each
      cluster.  Clusters are (top to bottom): high-activity, medium-high
      medium-low and low.}
    % \inote{Mauricio: change labels of x and y-axis (to: Event type
    % (total number of events))}
  }
  \label{fig:fig3}
\end{figure}

The collection of events is converted into their VQ-event model
representation. Using this model, we can identify events that have
produced similar levels of activity in the social network. In other
words, events are considered to have similar activity if the
interarrival times between their social media posts are similarly
distributed, implying a very much alike collective reaction from users
to the events within a group. In order to identify groups of similar
events, we cluster the event models. We sort the resulting groups of
events from highest to lowest activity, according to the concentration
of social media posts in the bins that correspond to short
interarrival times. We consider the events that fall in the top
cluster to be high-activity events as most of their interarrival times
are concentrated in the smallest interval of the VQ-event model.  In
our dataset, these correspond to roughly 8\% of the events.  We
consider the next clusters in the sorted ranking to form medium-high
activity events, and so on.  Thus we end with four groups of events:
high, medium-high, medium-low and low. Figure~\ref{fig:fig3} shows a
heatmap of the interarrival relative frequency for each cluster. This
classification of events based on activity intensity is independent of
event size. More details of this methodology are provided in the
supplementary material.

% Results and Discussion can be combined.
\section*{Results and Discussion}
%\newtext{ 
Our main objective in this work is to analyze the
  characteristics of high-activity events which differentiate them from
  other types of events. In particular, we identify how early on in an
  event's lifecycle can we determine if an event is going
  produce high activity in the on-line social network.

Tables~\ref{table:high-impact-sample}
and~\ref{table:low-impact-sample} show examples of events from the
high-activity category and low-activity category. We recall that the
high-activity events are those which were in the top 8\% of the
ranking obtained by sorting the event clusters according to
concentration of interarrival times of social media posts in the
shortest interarrival time of the VQ-event model.  Table
\ref{table:high-impact-sample} shows two events of different sizes
(large and small) and different scopes (one global and the other of
more local scope) categorized as high activity in our dataset. The
first event, the death of Nelson Mandela, is one of the largest events
in the dataset, with $\approx 134,000$ tweets. The histogram
representation of this event, shown in Figure \ref{fig:fig1a}, suggests
that more than $80\%$ of the activity of the event was produced in
high-activity periods.  This is an event of international, political,
and social importance, that produced an overwhelming flood of messages
on social media. %% add detail about reach and scope 
Hence, it makes
sense for such an example to be a high-activity event.  The second
event, on the other hand, about the 2013 Mumbai Gang Rape is of much
smaller scale, with a total of $\approx 1,700$ tweets.  However, this
event caused considerable amount of immediate reaction on social
media, with close to $50\%$ of its activity concentrated within
high-activity periods. Despite its smaller size, in comparison to the
previous event, this event displays a similar reaction to that of
other high-activity events, but at a smaller scale. 
%More detailed
%inspection of the event shows that geolocated users which discussed
%this event were mostly from India and the U.K. In addition, we find
%that in the following days other events that are repercussions of this
%event gather a great deal of attention (with $\approx 12,700$
%tweets). % is it good to include this fact??

Table \ref{table:low-impact-sample} shows events that have been
classified by our methodology in the category of low activity.  The
first event, about a teen surviving after hiding in the wheel of a
airplane, had only a little more than $25\%$ of its messages arriving
with high-activity bursts although it had over $18,000$ messages.  The
second event, about the damages caused by a tornado in Canada, did not
garner much immediacy in attention of Twitter users, with only $7\%$
of its messages produced with short interarrival times. Most of the
messages of this event were well spaced out in time. Even though we
cannot say whether or not this event had significant implications in
the real-world, we can say that it did not have considerable impact on
the Twitter network. The lack of interest could be due to several
factors that are currently beyond the scope of this work, ranging from
the lack of Twitter users in the locality of the real-world event, to
it not being considered urgent by Twitter users. We intend to research
the relation between the real-world impact of an event and the network
reaction in future work.

%%%%%%%%% REWRITE ANALYSIS OF HISTOGRAM DISTRIBUTIONS %%%%%%%%%%%%

% Fig.~\ref{fig:highest}, Fig.~\ref{fig:low}, Fig.~\ref{fig:cdf-highest} and
% Fig.~\ref{fig:cdf-lowest} show the average histograms and the average
% cumulative distribution functions of the events corresponding to the
% high and low activity categories, respectively.  Visually, the average
% high-activity event vector representations starkly differ from that of
% a low-activity event in that the histogram in Fig.~\ref{fig:highest}
% seems to possess an exponential decay, while the histogram in
% Fig.~\ref{fig:low} does not.  To test this hypothesis, we fit
% exponential function of the form $f(x)=ae^{bx}$ to the event
% histograms. Table~\ref{tab:curve_fitting} summarizes the results from
% statistical significance tests performed on the parameters $a$ and
% $b$, and on the residual least squares error used for fitting the
% exponential curves.  The differences between these values is
% statistically significant ($p$-value $\leq2.2\times 10^{-16}$), thus demonstrating
% that high-activity events, on an average, fit the exponential decay
% curve much better than their low-activity counterparts.  In addition,
% Fig.~\ref{fig:param_est} shows two scatter plots with the resulting
% exponential parameters $a$ and $b$.  We observe that the majority
% ($97.4\%$) of high-activity events have an exponent $b \leq -50$,
% separating them unequivocally from other events.

%}

\begin{figure}
  \centering
    \includegraphics[width=\textwidth]{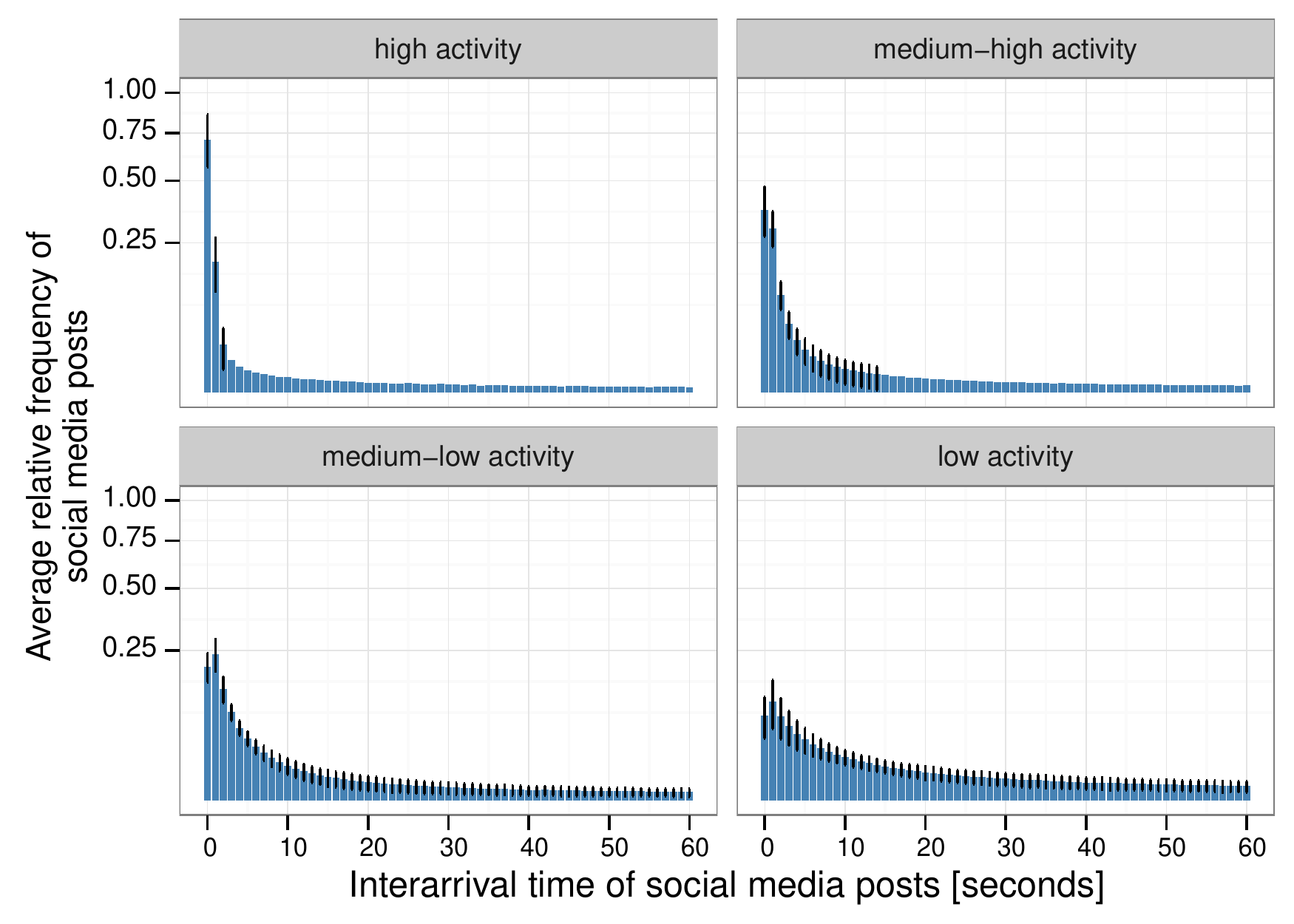}
  \caption{\textbf{Average histograms of the high activity,
      medium-high activity, medium-low activity and low activity
      clusters in our dataset (from left to right and top to
      bottom). All histograms include standard deviation bars and were
      cut-off at 60 second length for better visibility.
      % \inote{change labels of x and y axis}
    }}\label{fig:fig4}
\end{figure}

\begin{figure}
  \centering
    \includegraphics[width=\textwidth]{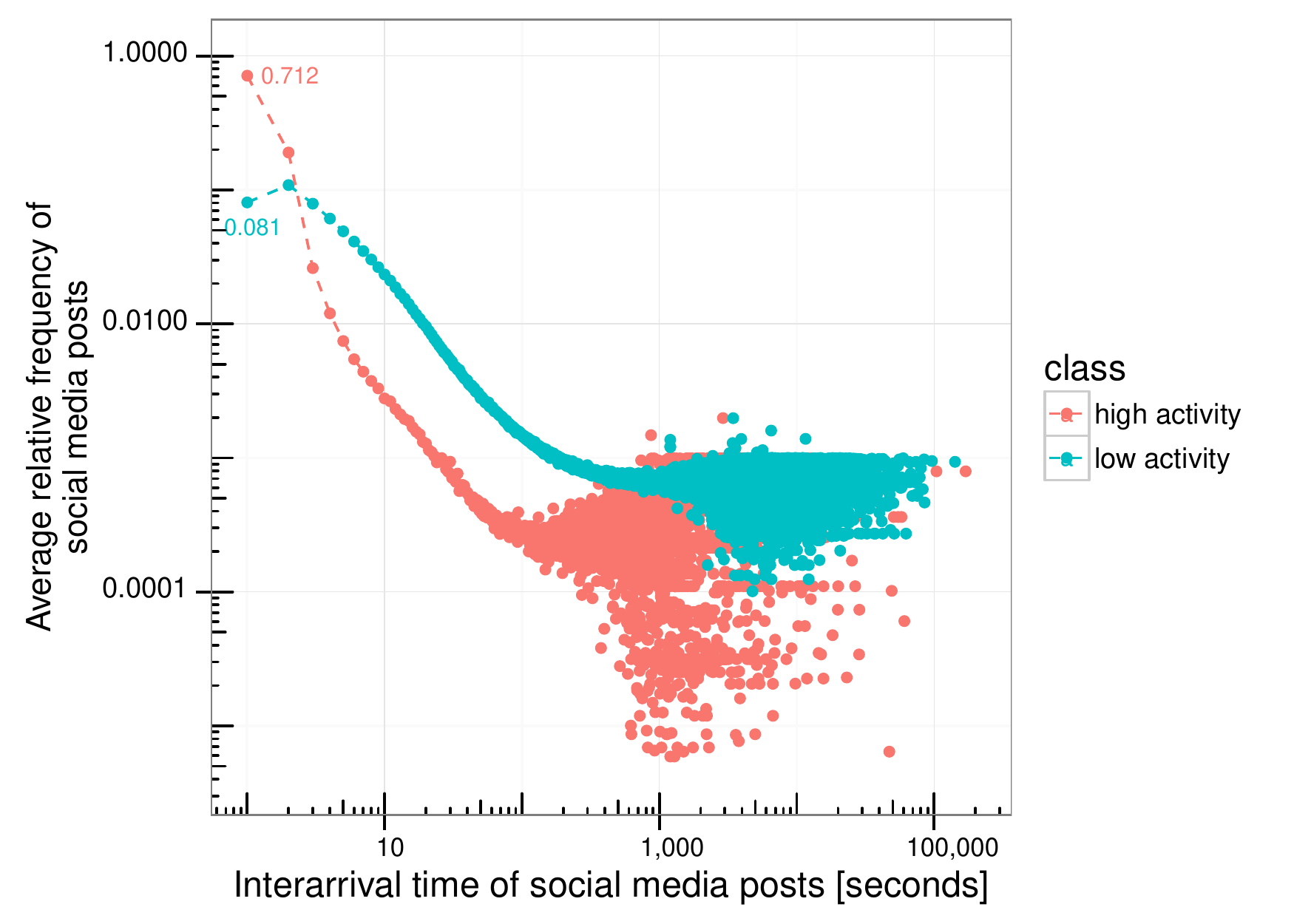}
  \caption{\textbf{Scatter plots of the average relative frequencies of interarrival times
for the high-activity and low-activity clusters of events (i.e., scatter
plots of the histograms in Fig.~\ref{fig:fig4} in log-log scale). $y$-axis
represents the average relative frequency of social media messages and
$x$-axis the interarrival time.
    }}\label{fig:fig5}
\end{figure}

\begin{figure}
  \centering
 \includegraphics[width=\textwidth]{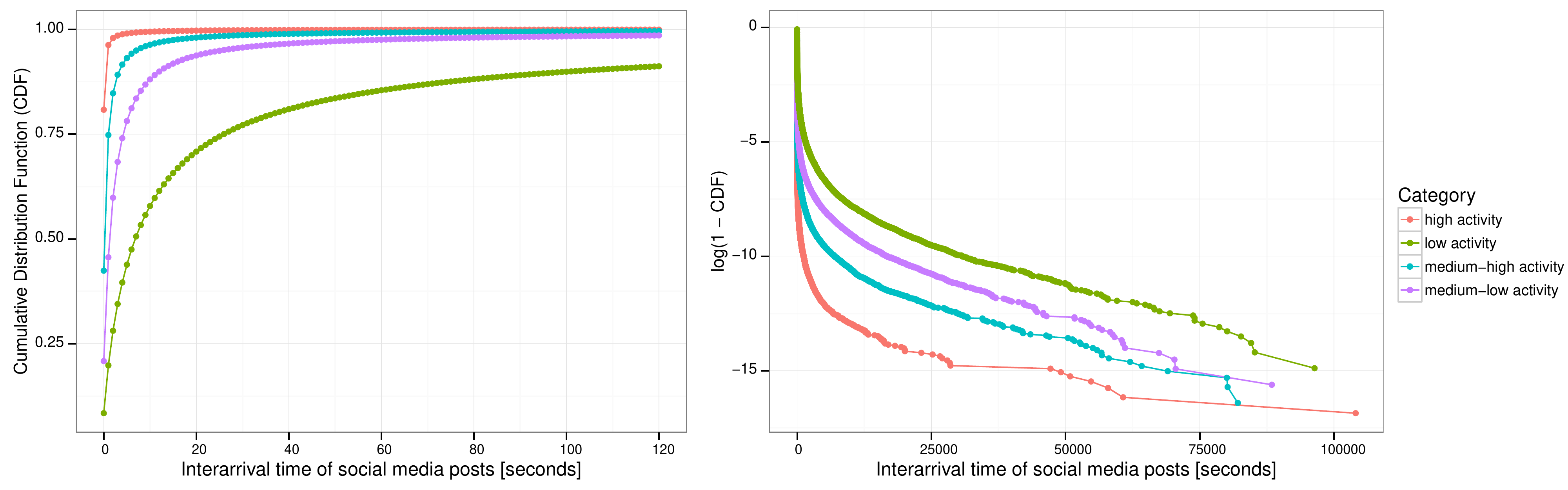}
  \caption{\textbf{(Left) Average cumulative distribution function (CDF) for
the high activity, medium-high activity, medium-low activity and low activity clusters in our dataset.
(Right) $\log{(1 - \mathrm{CDF})}$ for the same clusters. 
      % \inote{change labels of x and y axis}
    }}\label{fig:fig6} %% fig6_long_inf_omitted.pdf
\end{figure}

Fig.~\ref{fig:fig4} shows the average histograms for events that
belong to the high activity, medium-high activity, medium-low activity
and low-activity clusters (displayed from left to right and top to
bottom). All histograms show a quick decay in average relative
frequency (resembling a distribution from the exponential family). In
particular, the high-activity group concentrates most of its activity
in the shortest interarrival rate, with lower activity groups mostly
concentrating their activity in the second bin with slower
decay. Fig.~\ref{fig:fig5} further characterizes the differences in
behavior of the high and low-activity groups, showing that
high-activity events concentrate on average $70\%$ fo their activity
in the smallest bin ($0$ sec.), against $8\%$ for low-activity
events. In addition, Fig.~\ref{fig:fig6} (left) shows the cumulative
distribution function (CDF) for each group of events, and
Fig.~\ref{fig:fig6} (right) shows $\log{(1 - \mathrm{CDF})}$. Visual
inspection shows a clear difference in how interarrival rates are
distributed within each group, however, these figures do not indicate a
power-law distribution nor exponential distribution.%%

%%% Table with example tweets for events
%\clearpage
\begin{table}
  \centering
  {\scriptsize
    \begin{tabular*}{1\linewidth}{p{5cm}p{5cm}}
      \toprule
      \textbf{Event} & \textbf{Sample Tweets} \\
      \midrule
      \pbox{20cm}{\textbf{Description:}\\ Death of South African\\ politician Nelson Mandela. \vspace{.1cm}\\
        \textbf{Keywords:}\\ {[}nelson, mandela{]}\vspace{.1cm}\\
        \textbf{Date:}\\ 2013-12-05 \vspace{.1cm}\\
        \textbf{Size:} \\ 134,637 tweets}
      & \pbox{20cm}{
        @DaniellePeazer: RIP Nelson Mandela..... what a truly phenomenal and\\ inspirational man xx\vspace{.1cm}\\
        @iansomerhalder: Im in tears.The world has lost one of its greatest shepherds \\of peace. Thank you Mr.Mandela for the love you radiated. http://t.co/u39MVVEKe8\vspace{.1cm}\\
        @FootballFunnys: This is so true. RIP Nelson Mandela. http://t.co/vF9xri8LdP\vspace{.1cm}\\
        @David\_Cameron: I've spoken to the Speaker and there will be statements \\and tributes to Nelson Mandela in the House on Monday.} \\
      \midrule
      \pbox{20cm}{\textbf{Description:}\\ 2013 Mumbai Gang Rape \vspace{.1cm}\\
        \textbf{Keywords:}\\ {[}rape, mumbai{]}\vspace{.1cm}\\
        \textbf{Date:}\\ 2013-08-24 \vspace{.1cm}\\
        \textbf{Size:} \\1,705 tweets}
      & \pbox{20cm}{
      % select user.screen_name, text from tweet join user on tweet.user_id_id = user.user_id where event_id_id = 272;
      @TheNewsRoundup: Mumbai gang-rape: Second accused confesses to crime: \\Mumbai Police - Daily News Analysis http://t.co/KnabwhqH66\vspace{.1cm}\\
      @vijayarumugam: An interesting take on the Mumbai rape: http://t.co/ylBmW4l8sA\vspace{.1cm}\\
      @LondonStephanie: Two arrested over gang rape of Mumbai photojournalist \\that sparked renewed protests in India http://t.co/McYfLNDvaE\vspace{.1cm}\\
      @GanapathyI: Most brutal rapist of Delhi gang-rape was 17. Most brutal rapist\\ of Mumbai gang-rape is 18. Worst Young generation I have seen in my life.}\\
%        @M\_arioBalotelli: CLEAR ANGLE of the Suarez bite!!  https://t.co/bI08YsZWSE\vspace{.1cm}\\
%        @fifamedia: Disciplinary proceedings opened against Uruguay's Luis Suarez\\ http://t.co/w6mRNuSGZt\vspace{.1cm}\\
%        @DeadlineDayLive: Luis Suarez will sign a 5-year contract at Barcelona and he'll wear \\the no. 9 shirt. (Source: http://t.co/6uRIUwjsGN) http://t.co/FxyOf9ERVr\vspace{.1cm}\\
%        @GeniusFootball: BREAKING: FIFA have caught Suarez leaving the stadium?\\ http://t.co/vsQQCVV1GQ} \\
      \bottomrule
    \end{tabular*}
  }
  \caption{\textbf{
        Examples of high-activity news events. The events
      shown were taken from the ``high'' category according to
Fig.~\ref{fig:fig4}.
        }}
  \label{table:high-impact-sample}
\end{table}

\begin{table}
  \centering
  {\scriptsize
    \begin{tabular*}{1\linewidth}{p{5cm}p{5cm}}
      \toprule
      \textbf{Event} & \textbf{Sample Tweets} \\
      \midrule
      \pbox{20cm}{\textbf{Description:}\\Teen survives hiding \\in a plane wheel.\vspace{.1cm}\\
        \textbf{Keywords:}\\ {[}teen, survives, old, \\well, skydivers, plane, wheel, flight{]}\vspace{.1cm}\\
        \textbf{Date:}\\ 2014-04-21 \vspace{.1cm}\\
        \textbf{Size:}\\ 18,519}
      & \pbox{20cm}{
        %select user.screen_name, text from tweet join user on tweet.user_id_id = user.user_id where event_id_id in (22310,22274,22240);
        @ToniWoemmel: 16-year-old somehow survives flight from California to\\ Hawaii stowed away in planes wheel well: http://t.co/IGiJa60SiK\vspace{.1cm}\\
        @iOver\_think: 38,000 feet at -80F: Teen stowaway survives five-hour\\ California-to-Hawaii flight in wheel well http://t.co/ejXQH9VZyT\vspace{.1cm}\\
        @TruEntModels: GOD IS GOOD...runaway TEEN hid in plane's wheel for\\ 5 HOUR flight during FREEZING temps and survived http://t.co/6g6Cqhs9Ib\vspace{.1cm}\\
        @DvdVill: A 16-year-old kid, who was mad at his parents, hid inside a jet\\ wheel and survived flight to Hawaii. http://t.co/c82GbjrfUH\\
        %@guardiannews: Angela Merkel denied access to her NSA file http://t.co/FLQc0zSjYJ\vspace{.1cm}\\
        %@mog7546: \#GERMANY's \#Merkel says \#OBAMA's \#US assurances on \#NSA spying\\ "INSUFFICIENT" - Reuters India http://t.co/D2L52CP9YZ\vspace{.1cm}\\
        %@GermanyForum: Merkel denied access to own NSA file http://t.co/e6vKCOkbXA\vspace{.1cm}\\
        %@kgosztola: US ignores request from German Chancellor Angela \\Merkel to look at NSA file:\\ http://t.co/HFTMMZBu5W
      }
      \\
            \midrule
      \pbox{20cm}{\textbf{Description:}\\Surveying the damages of \\ recent tornado in Canada. \vspace{.1cm}\\
        \textbf{Keywords:}\\ {[}canada, tornado{]}\vspace{.1cm}\\
        \textbf{Date:}\\ 2014-06-21 \vspace{.1cm}\\
        \textbf{Size:}\\ 1,033}
      & \pbox{20cm}{
        @Kathleen\_Wynne: Visited \#Angus today to survey the damage. Thankfully no \\fatalities or major injuries from recent tornado. http://t.co/xRQyRWg5Vw\vspace{.1cm}\\
        @SunNewsNetwork: PHOTOS \& VIDEO: Hundreds displaced after \\ tornado hits Ontario town, destroying homes http://t.co/L38rG6N1a6\vspace{.1cm}\\
        @CBCToronto: Kathleen Wynne is speaking at site of tornado damage in Angus, \\Ont. now. Watch live here: http://t.co/EDKNUiZo0X \#cbcto\vspace{.1cm}\\
        @InsuranceBureau: @CTVBarrieNews: Insurance Bureau of Canada is setting up \\a mobile unit in \#Angus today to help residents affected by \#Tornado}
      \\
      \bottomrule
    \end{tabular*}
  }
  \caption{\textbf{
                Examples of events with low activity. The events
      shown were taken from the ``low'' category according to
Fig.~\ref{fig:fig4}}
          .}
  \label{table:low-impact-sample}
\end{table}

Further analysis of the high-activity events shows significant
differences to other events, in the following aspects: (i) how the
information about these events is propagated, (ii) the characteristics
of the conversations that they generate, and (iii) how focused users
are on the news topic. In detail, high-activity events have a higher
fraction of {\em retweets} (or shares) relative to their overall
message volume. On average, a tweet from a high-activity event is
retweeted 2.36 times more than a tweet from a low activity event. The
most retweeted message in high-activity events is retweeted 7 times more
than the most retweeted message in a medium or low activity event. We
find that a small set of initial social media posts are propagated
quickly and extensively through the network without any rephrasing by
the user (just plain forwarding). Intuitively, this seems justified given
general topic urgency of high-activity events. Events that are not
high-activity did not exhibit these characteristics.

Our research also revealed that high-activity events tend to spark more conversation
between users, 33.4\% more than other events. This is reflected in the
number of {\em replies} to social media posts. The number of different
users that engage with high-activity events is 32.7\% higher than in
events that are not high-activity. Posts about high-activity events are
much more topic focused than in other events. The vocabulary of unique
words as well as {\em hashtags} used in high-activity events is much
more narrow than for other events. Medium and low activity events have
over 7 times more unique hashtags than high-activity events. This is
intuitive, given that if a news item is sensational, people will
seldom deviate from the main conversation topic.

% We have presented an analysis of high-impact news events based on
% the data of their entire life-cycle in the social network. We used
% the arrival time intervals to create a model that allows us to
% classify the event according to its impact. Nevertheless,

In a real-world scenario, in order to predict if an early breaking
news story will have a considerable impact in the social network, we
will not have enough data to create its activity-based model, i.e., we
will not yet know the distribution of the speed at which the social
media posts will arrive for the event. For instance, an event can
start slowly and later produce an explosive reaction, or start
explosively and decay quickly to an overall slower message arrival
rate. Still, reliable early prediction of very high-activity news is
important in many aspects, from decisions of mass media information
coverage, to natural disaster management, brand and political image
monitoring, and so on.

For the task of early prediction of high-activity events we use features 
that are independent of our activity-based model such as
the retweets, the sentiment of the posts about the event, etc. 
These features are computed on the early 5\% of messages about the event.
% More details are provided in the supplementary material.
The results are an average from a 5-fold cross validation with
randomly selected 60\% training, 20\% validation and 20\% test splits.
The high-activity events are identified with a precision of 82\% using
only the earliest 5\% of the data of each event
(Table~\ref{tab:classification_results}).  Additionally, we were able to
identify with high accuracy a considerable percentage of all
high-activity events ($\approx 46\%$) at an early stage, with very few
false positives (Table~\ref{tab:classification_results} and~\ref{tab:confusion_matrix}).

\begin{table}
  %\begin{adjustwidth}{-10mm}{-10mm}
  \centering
  {\small
    \begin{tabularx}{\textwidth}{lcccc|cccc}
      \toprule
      & \multicolumn{4}{c}{\textbf{Early 5\% Tweets}} & \multicolumn{4}{c}{\textbf{All Tweets}} \\
      \midrule
      & FP-Rate & Precision & Recall & ROC-area & FP-Rate & Precision & Recall & ROC-area \\
      % \midrule
      high-activity & 0.009 & 0.819 & 0.455 & 0.900 & 0.01 & 0.830 & 0.540 & 0.945 \\
      non-high-activity & 0.545 & 0.954 & 0.991 & 0.900 &  0.460 & 0.960 & 0.990 & 0.945 \\
      \bottomrule
    \end{tabularx}
  }
  \caption{\textbf{Classification of high-activity events.}}
  \label{tab:classification_results}
  %\end{adjustwidth}
%                                                                                                                                448,1         93%
\end{table}

\begin{table}
  \centering
  % {\scriptsize
  \begin{tabularx}{\textwidth}{lcc|cc}
    \toprule
    \multirow{2}{*}{ }& \multicolumn{2}{c}{\textbf{Early 5\% Tweets}} & \multicolumn{2}{c}{\textbf{All Tweets}} \\
    \midrule
    % \cmidrule{2-5} \cline{2-5}
    & high-activity & non-high-activity & high-activity & non-high-activity \\
    % \midrule
    high-activity & $194$ & $232$ & $230$ & $196$\\
    non-high-activity & $43$ & 4,765 & 47 & 4,761 \\
    \bottomrule
  \end{tabularx}
  % }
  \caption{\textbf{Confusion matrix for high-activity events prediction.}}
  \label{tab:confusion_matrix}
\end{table}

The precision using only the early tweets is almost as good as using
all tweets in the event (0.819 to 0.830). This suggests that the
social network somehow acts as a natural filter in separating out the
high-activity events fairly early on.  The recall goes from 0.455 to
0.540. This indicates that there are some high-activity events which
require more data in order to determine what kind of activity they will
produce, or events for which activity occurs due to random conditions. A
detailed description of the features and different classification
settings are provided in the supplementary material.%\supplementary.

\section*{Conclusion}

We study the characteristics of the activity that real-world news produces
in the Twitter social network. In particular, we propose to measure the impact of the
real-world news event on the on-line social network by modeling the user
activity related to the event using the distribution of their
interarrival times between consecutive messages. In our research we observe
that the activity triggered by real-world news events follows a similar
pattern to that observed in other types of collective reactions to events.
This is, by displaying periods of intense activity as well as long periods of
inactivity. We further extend this analysis by identifying groups of events
that produce much more concentration of high-activity than other events. 
We show that there are several specific properties that distinguish how
high-activity events evolve in Twitter, when comparing them to other
events. We design a model for events, based on the codebook approach, that allows us to do
unambiguous classification of high-activity events based on the impact
displayed by social network. % This definition does not have some of the
%problems that current notions of virality and popularity have. 
Some notable
characteristics of high-activity events are that they are forwarded more
often by users, and generate a greater amount of conversation than
other events.  Social media posts from high-activity news events are
much more focused on the news topic. Our experiments show that there
are several properties that can suggest early on if an event will have
high-activity on the on-line community.  We can predict a high number of
high-activity events {\em before} the network has shown any type of
explosive reaction to them. % Using simple off-the-shelf feature based
%classifiers, we can
% predict many high-impact events with high precision.
This suggests that users are collectively quick at deciding whether an
event should receive priority or not.  However, there does exist a fraction of
events which will create high activity, despite not presenting
patterns of other high activity events during their early stages.  These
events are likely to be affected by other factors, such as random
conditions found in the social network at the moment and require
further investigation.
\nolinenumbers

\bibliographystyle{plain}
\bibliography{refs}

\section*{Supporting Information}
\subsection*{S1 Appendix}
\section{Data Collection Methodology}
\label{sec:data_collection_methodology}
The Twitter Search
API\footnote{\url{http://dev.twitter.com} (Accessed: August 25, 2015)} was used to obtained tweets
about $5,234$ news events.  
This encompasses a total of
$43,256,261$ tweets. Table~\ref{table:dataset-stats} shows a high level
description of the dataset.  The full
  dataset is available in
  \url{http://dcc.uchile.cl/~mquezada/breakingnews/}.

% Please add the following required packages to your document
% preamble: \usepackage{booktabs}
\begin{table}[h]
  \centering
  \begin{tabular}{@{}lllll@{}}
    \toprule
    \textbf{News events' property} & \textbf{Minimum} & \textbf{Mean} & \textbf{Median} & \textbf{Maximum} \\ \midrule
    \# of tweets & $1,000$ & $8,254$ & $2,474$ & $510,920$ \\
    \# of keywords & $2$ & $3.77$ & $3$ & $39$ \\ 
    Event duration (hours) & $0.12$ & $20.93$ & $7.46$ & $190.43$ \\ \bottomrule
  \end{tabular}
  \caption{\bf High-level description of the dataset of news events.} \label{table:dataset-stats}

\end{table}

\subsection{Collecting the Tweets}
\label{sec:collecting_the_data}
\begin{figure}
  \includegraphics[width=\textwidth]{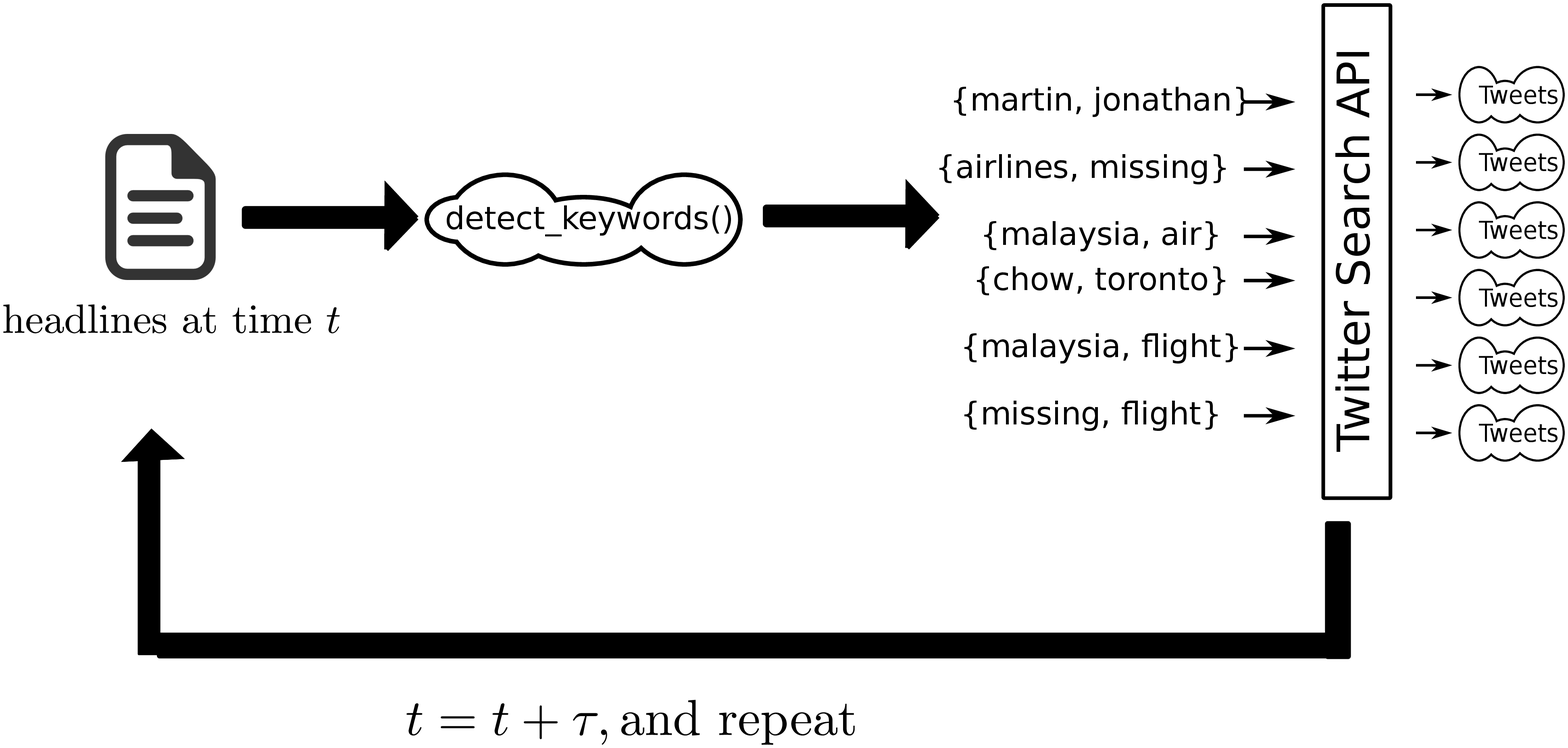}
  \caption{\textbf{This figure illustrates the high level data
      collection process. Headlines are collected every hour, and $6$
      keyword pairs are chosen to search for tweets. These keyword
      pairs are detected with the goal of concisely representing
      queries for an event.}}
  \label{fig:data_collection_1}
\end{figure}

The data collection process entails detecting pairs
of keywords from the most recent hourly batch of news headlines (the
pairs of keywords are meant to describe the events succinctly), and
then searching for tweets using the pairs of keywords as queries.
We merge the search results of
`similar' queries every 24 hours and form the tweets set for an event.
We obtained the hourly batch of headlines from the news media accounts
on Twitter listed in Table~\ref{table:acct_names}.
Figure~\ref{fig:data_collection_1} represents the high-level flowchart
of the data collection process. A summary of this process is described
in Algorithm~\ref{alg:data_collection}.
The accounts are verified accounts on
Twitter\footnote{Verified accounts on Twitter establish authenticity
  of identity of key individuals and organizations.}.

{\footnotesize
  \begin{longtable}{lll}
    \toprule
    \textbf{Twitter Account} & \textbf{Name} & \textbf{Location} \\
    \midrule
    \endfirsthead
    \multicolumn{3}{l}%
    {\tablename\ \thetable\ -- \textit{Continued from previous page}} \\
    \toprule
    \textbf{Twitter Account} & \textbf{Name} & \textbf{Location} \\
    \midrule
    \endhead
    \hline \multicolumn{3}{l}{\textit{Continued on next page}} \\
    \endfoot
    \bottomrule
    \endlastfoot

    breakingnews     &  Breaking News         &  Global                      \\
    cnnbrk           &  CNN Breaking News     &  Everywhere                  \\
    cnn              &  CNN                   &                              \\
    nytimes          &  The New York Times    &  New York City               \\
    bbcbreaking      &  BBC Breaking News     &  London, UK                  \\
    theeconomist     &  The Economist         &  London                      \\
    skynewsbreak     &  Sky News Newsdesk     &  London, UK                  \\
    reuters          &  Reuters Top News      &  Around the world            \\
    wsjbreakingnews  &  WSJ Breaking News     &  New York, NY                \\
    foxnews          &  Fox News              &  U.S.A.                      \\
    msnbc\_breaking   &  msnbc.com Breaking    &                              \\
    skynews          &  Sky News              &  London, UK                  \\
    nbcnews          &  NBC News              &  New York, NY                \\
    cbsnews          &  CBS News              &  New York, NY                \\
    bbcworld         &  BBC News (World)      &  London, UK                  \\
    abc              &  ABC News              &  New York, NY                \\
    bbcnews          &  BBC News (UK)         &  London                      \\
    ap               &  The Associated Press  &  Global                      \\
    telegraphnews    &  Telegraph News        &  London, UK                  \\
    breakingnewsuk   &  Breaking News UK      &  London                      \\
    channel4news     &  Channel 4 News        &  Weekdays at 7 on Channel 4  \\
    twcbreaking      &  TWC Breaking          &  Atlanta, GA                 \\
    washingtonpost   &  Washington Post       &  Washington, D.C.            \\
    yahoonews        &  Yahoo News            &  Santa Monica, Calif.        \\
    breakingpol      &  Breaking Politics     &  Global                      \\
    nydailynews      &  New York Daily News   &  New York City               \\
    ajenglish        &  Al Jazeera English    &  Doha, Qatar                 \\
    usatoday         &  USA TODAY             &  USA TODAY HQ, McLean, Va.   \\
    wsj              &  Wall Street Journal   &  New York, NY                \\
    guardiannews     &  Guardian news         &  London                      \\
    bloombergnews    &  Bloomberg News        &  New York and the World      \\
    abcworldnews     &  ABC World News        &  New York                    \\
    nypost           &  New York Post         &  New York, NY                \\
    msnbc            &  msnbc                 &                              \\
    nbcnightlynews   &  NBC Nightly News      &  New York                    \\
    huffingtonpost   &  Huffington Post       &                              \\
    rt\_com           &  RT                    &                              \\
    abcnews          &  ABC News              &  Australia                   \\
    latimes          &  Los Angeles Times     &  Los Angeles, CA             \\
    googlenews       &  Google News           &  Mountain View, CA           \\
    cnnlive          &  CNN Live              &  Everywhere                  \\
    newshour         &  NewsHour              &  Arlington, VA               \\
    guardian         &  The Guardian          &  London                      \\
    afp              &  Agence France-Presse  &  France                      \\
    independent      &  The Independent       &  London, United Kingdom      \\
    ndtv             &  NDTV                  &  India                       \\
    cp24             &  CP24                  &  Toronto                     \\
    reuterslive      &  Reuters Live          &  Global                      \\
    bostonglobe      &  The Boston Globe      &  Boston, MA                  \\
    foxnewsalert     &  Fox News Alert        &  New York, NY                \\
    ft               &  Financial Times       &  London                      \\
    jerusalem\_post   &  The Jerusalem Post    &  Israel                      \\
    bbcnewsus        &  BBC News US           &  Washington DC               \\
    foxheadlines     &  Fox News              &  New York, NY                \\
    forbes           &  Forbes                &  New York, NY                \\
    thetimes         &  The Times of London   &  London                      \\
    usnews           &  U.S. News             &  Washington, DC\\

    \caption[List of news accounts.]{\textbf{List of news account. The
        first column is the Twitter account. It can be accessed in a
        browser at \texttt{http://twitter.com/accountname}. The second
        and third columns were obtained from each account's page.}}
    \label{table:acct_names}
\end{longtable}
}

In Algorithm~\ref{alg:data_collection}, the goal of the {\tt
  detect\_keywords()} module is to produce pairs of keywords that
coherently, and succinctly describes an event. Inspired by the data
mining concept of mining frequent itemsets \cite{Tan_Steinbach_Kumar},
we develop an algorithm which identifies the most commonly occurring
keyword groups (or item sets) in the headlines. From the item sets, we
pick the most common keyword pairs. The algorithm is described in
Algorithm~\ref{alg:detect_keywords}. This algorithm finds string
intersections between headlines ({\tt intersect()} in
Line~\ref{alg:line:intersect} returns the number of words present in
both $s_a$ and $s_b$). If the common set of words has sufficient
Jaccard similarity to any of the existing item sets, then the common
set of words are added to that item set. If not, a new item set is
created (Line~\ref{alg:line:create}). During the process of
identifying the most commonly occurring item sets, we also track how
many times each keyword has been added to an item set, namely, the
score of the keyword. The score of each item set is the average of the
scores of its
keywords. %\footnote{The module {\tt addOne() adds 1 to the score of
% every keyword in the input set.}}.
Once the item sets have been identified, we select the top 2 keywords
from each of the top six item sets and use them for searches. We
preprocess the headlines to remove duplicates, stopwords, punctuation,
convert everything to lower case, and subject the text through the
process of stemming.

We made the choice of selecting 2 keywords since having a single
keyword maybe not define an event accurately. For example, the keyword
\{obama\} could retrieve tweets about any event related to Obama.
However, a keyword \emph{pair} like \{obama, syria\} describes the
event more accurately\footnote{Having more than two keywords may
  impose too much of a restriction on the query, leading to little or
  no tweets in the retrieval.}.

The Twitter Search API imposes several restrictions on the number of
searches that can be performed in a given time duration.
% In order to circumvent this restriction and capture as many tweets
% as we possibly can,
We produce six search threads to perform searches, one for each
keyword pair. All in all, with $\tau = 60$ minutes in
Figure~\ref{fig:data_collection_1}, six new pairs of keywords are
discovered from the most recent batch of headlines, and then we query
for tweets in the Twitter Search API using these keywords over the
next one hour.

We make some notes about the data collection methodology. Firstly,
there is a temporal sensitivity to the data collection methodology.
For example, one of the keyword pairs obtained as soon the Malaysian
airlines jet disappeared was \{plane,missing\}. Although this keyword
pair does not specifically refer to the Malaysian airlines jet, it is
likely that the tweets retrieved from searching for this pair will
indeed be about the Malaysian airlines plane that went missing, since
the search is performed as and when the event breaks out. Secondly,
Algorithm~\ref{alg:detect_keywords} may return multiple pairs of
keywords (possibly different pairs) describing the same event. Some
pair examples of keywords produced when there was a bomb threat at
Harvard University in December 2013 were \{harvard, evacuated\},
\{harvard, explosives\}, etc. How do we merge the keyword pairs which
belong to the same event? In order to address this, we collect all the
pairs obtained in the past $24$ hours, and build a graph with keywords
as nodes, and keyword pairs (as obtained from Algorithm
{\ref{alg:detect_keywords}}) as edges. We then discover the connected
components of this graph, and treat each connected component as an
``event"\footnote{For the rest of the document, the terms
  \emph{connected component} and \emph{event} are used
  interchangeably. Both of them refer to the definition of
  \emph{event} given in the main article.}. The set of tweets obtained
by merging the tweets from each of the keyword pairs is the set of
messages associated with the event. Figure
\ref{fig:connected_components} is an example component formed on
December 16, 2013. It illustrates the merge of smaller keyword pairs
into larger components for two events. One was the bomb threat at
Harvard University, and the other was about the attack on police in
the Xinjiang province in China.

\begin{algorithm}
  \caption{{\tt data\_collection()}}
  \label{alg:data_collection}
  \begin{algorithmic}[1]
    \REQUIRE stream of headlines. \\
    \ENSURE data structures $\{\H_1, \H_2, \ldots \}$, with $\H.keywords$ = keyword pair, and $\H.tweets$ = set of tweets\\
    \STATE $i \leftarrow 0$, $j \leftarrow 0$ \\
    \LOOP
    \STATE $\mathcal{S} \leftarrow$ headlines for hour-$i$ \\
    \STATE $keyPairs \leftarrow$ {\tt detect\_keywords$(\mathcal{S})$}
    \COMMENT{$keyPairs$ is a list of keyword pairs.} \FOR{$k = 0$ to
      {\tt len($keyPairs$)$-1$}}
    \STATE $\H_j.keywords \leftarrow keyPairs[k]$ \\
    \STATE $\H_j.tweets \leftarrow $search$(\H_j.keywords)$
    \COMMENT{using Twitter Search API} \STATE $j \leftarrow j+1$
    \ENDFOR
    \STATE $i \leftarrow i+1$
    \ENDLOOP
  \end{algorithmic}
\end{algorithm}

% \begin{algorithm}
%   \caption{{\tt detect\_keywords()}}
%   \label{alg:detect_keywords}
%   \begin{algorithmic}[1]
%     \REQUIRE A set of headlines, $\ess = \{s_1, s_2, \ldots\}$. \\
%     \ENSURE Up to $6$ pairs of keywords, $\{k1,k2\}_{j=1}^{6}$
%     \STATE $\I_i \leftarrow \emptyset$ for $i = 0,1,2, \ldots $
%     \COMMENT{\# Initialize item sets to null.} \FOR{$\forall$
%     $\{a,b\}$ such that $\{s_a, s_b\} \in \ess$} \STATE $\G
%     \leftarrow ${\tt intersect($s_a,
%     s_b$)} \label{alg:line:intersect} \IF{$|\G| \geq \eta$} \STATE
%     $noMatches \leftarrow 1$ \FOR{$i = 0,1,2, \ldots$} \IF{$|${\tt
%     intersect($\G, \I_i$)} $| \geq \gamma$} \STATE $\I_i \leftarrow
%     ${\tt union($\G,\I_i$)} \STATE {\tt addOne($\I_{i}$)} \STATE
%     $noMatches = 0$
%     \ENDIF
%     \ENDFOR
%     \IF{$noMatches$} \STATE Create $\I_{i+1} \leftarrow
%     \G$ \label{alg:line:create} \STATE {\tt addOne($\I_{i+1}$)}
%     \ENDIF
%     \ENDIF
%     \ENDFOR
%     \STATE $ keyPairs = $ {\tt{<empty list>}} \FOR{$\I \in$ {\tt
%     sorted($\I_1, \I_2, \ldots$)[:6]}} \STATE $keyPairs.append$({\tt
%     sorted($\I$)[:2]})
%     \ENDFOR
%     \STATE return $keyPairs$
%   \end{algorithmic}
% \end{algorithm}

% \renewcommand{\algorithmicrequire}{\textbf{Input:}}
% \renewcommand{\algorithmicensure}{\textbf{Output:}}

\begin{algorithm}
  \caption{\tt{detect\_keywords()}}
  \label{alg:detect_keywords}
  \begin{algorithmic}[1]
    \REQUIRE A set of $M$ sets of words, $\ess= \{H_1,H_2, \ldots,
    H_M\}$, positive integers $k, \eta$ \ENSURE $k$ sets of keywords,
    $G = (\I_1,\I_2,\ldots,\I_{k})$ \STATE $\I_i \leftarrow \emptyset$
    for $i = 1,2,
    \ldots,k$ %\COMMENT{Initialize item sets to empty sets}
    \STATE $score_i \leftarrow$ empty dictionary for $i = 1, 2,
    \ldots,k$ %\COMMENT{Initialize keyword set scores to $1$}
    \STATE $i \leftarrow 1$ \FOR{every pair of headlines $\{H_a, H_b\}
      \in \ess$ such that $|H_a \cap H_b| \geq \eta$} \STATE $\G
    \leftarrow H_a \cap H_b$
    \label{alg:line:intersect} %\COMMENT{$\G$ is the set of common words of $H_a$ and $H_b$}
    \STATE $j \leftarrow \operatorname{arg\,max}_j |\I_j \cap \G|$
    \IF{$|\I_j \cap \G| \geq \eta$} \STATE $\I_j \leftarrow \I_j \cap
    \G$ \STATE $score_{j}[w] \leftarrow score_{j}[w] + 1$ for all $w
    \in \I_j$ \ELSE \STATE $\I_i \leftarrow
    \G$ \label{alg:line:create} \STATE $score_{i}[w] \leftarrow 1$ for
    all $w \in \I_i$ \STATE $i \leftarrow i + 1$
    \ENDIF
    \ENDFOR
    \STATE $total\_score_i \leftarrow \sum_{w \in \I_i} score_{i}[w]$
    for $i = 1,2,\ldots,k$ \RETURN $G \leftarrow (\I_i$ sorted by
    $total\_score_i)$
  \end{algorithmic}
\end{algorithm}

\subsection{Cleaning the Data}
The data was preprocessed to reduce the noisy and irrelevant tweets.

\subsubsection{Special Stopwords:  Articulation Words}
During the data collection process, sometimes
unrelated events were joined together with keywords that was common to
both events.  

Typical stopwords such as ``the" and ``a" were removed during preprocessing
the news headlines. However, there are other words
which occur quite commonly in news headlines. For example, words like
``watch'', ``live'', or ``update'' are common to express things like
``watch this video'', ``we are live on TV'', or to update a previous
headline with more information. Such words could possibly incorrectly
connect two or more very different events as one.
Example: ``Watch Jim Harbaugh's press conference
live''\footnote{\url{https://twitter.com/49ers/status/519202023628374016}
(Accessed: August 25, 2015)}
and ``WATCH LIVE: Of the 48 people being monitored for contact with
Dallas patient, no one is showing any
symptoms''\footnote{\url{https://twitter.com/PzFeed/status/519203692898435072}
(Accessed: August 25, 2015)}.  We call such words
\emph{articulation words} 
We now delve into
understanding how and when these words occur, and how to subsequently
identify and remove them in the preprocessing step, just as we would a
stopword.

It is well known that \emph{tf-idf} \cite{Jones72astatistical} is a statistic of a word that
indicates how important that word is in a given document.  Intuitively, if a word appears
in all the documents, then its statistic is generally low in all the documents.  However,
if the word appears in very few documents, its statistic in those documents is fairly high,
indicating that the word is somehow representative of the content of the document.
It turns out the articulation words do not occur often enough for them to be detected by
regular \emph{tf-idf}, but do occur enough
times for them to falsely relate several unrelated events together. To
identify a group of those keywords, we used a modified \emph{tf-idf}
to detect them from the headlines.

The modified version of \emph{tf-idf}, what we refer as
\emph{maxtf-idf}, is meant to assign more weight to the terms that are
frequent in any document. For instance, \emph{tf-idf} of a term in a
document tries to assign a weight related to how ``rare'' that term is
in the whole collection, and how frequent the term is in that document,
thus indicating how representative the term is of the document. On the
other hand, we want to place a higher weight on a term if its frequency
is higher in any other document, relative to the frequency in the current
document. With that in mind, we want to identify terms that might be
``adding noise'' to the corpus and hence merge unrelated
events together.

The definition of \emph{maxtf} is as follows:

\begin{equation}
  \text{maxtf}(t,d,D) = 0.5 + \frac{0.5 + max\{f(t,d') : d' \in D\}}{max\{f(w,d) : w \in d\}}
\end{equation}

and for \emph{idf}, the usual formula:

\begin{equation}
  \text{idf}(t,D) = \log\frac{N}{|\{d \in D : t \in d\}|}
\end{equation}

where $t$ is a term, $d$ is a document, and $D$ is the corpus of all
documents. In this case, we set $t$ as a keyword, $d$ as the set of
keywords of one hour of a given day, and $D$ the set of documents of
that day.

\begin{figure}
  \begin{center}
    \includegraphics[width=\textwidth]{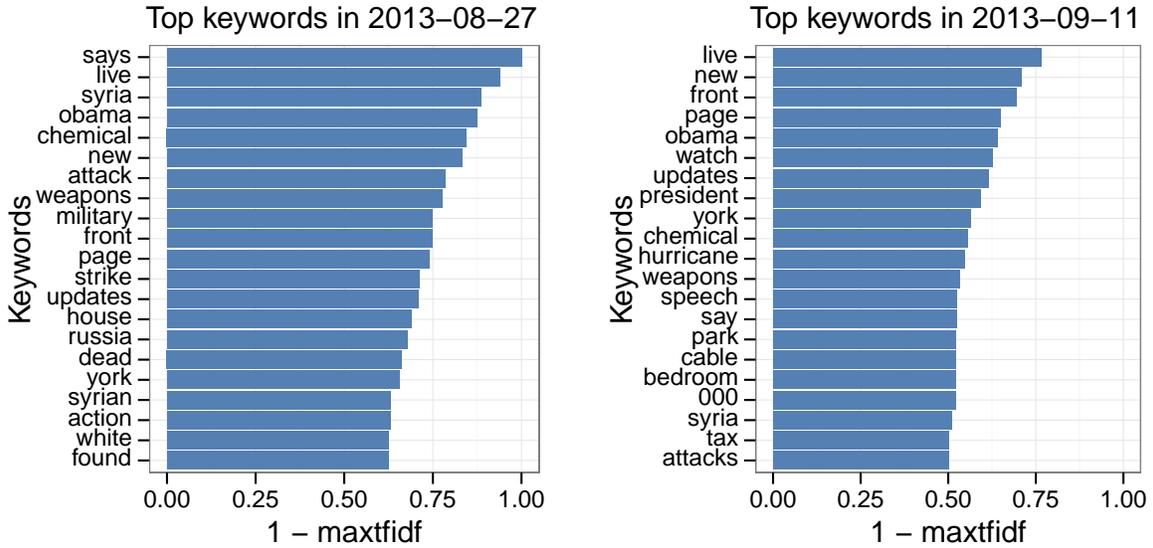}
    \caption[Stopwords detection.]{\textbf{Stopwords detection.
        Normalized $1-\text{maxtf-idf}$ score for data from August
        27th (left) and August 28th (right) of 2013. The top score
        words for both plots are ``says'' and ``live''. We used the
        top score words to disconnect connected components of
        events.}}
    \label{fig:stopwords}
  \end{center}
\end{figure}

After identifying such words, the idea is to disconnect the
components connected by those words. The process is to disconnect
each component by the word with top normalized $1-\text{maxtf-idf}$
score each time until the component could not be disconnected further.
We add the top scoring words to our list of stopwords.
These words are hence ignored from the subsequent runs of the data collection methodology.
In Figure~\ref{fig:stopwords} there are two examples of this process
to identify the words.
\subsubsection{Discarding Irrelevant Tweets}
\label{subsubsec:discarding_irrelevant_tweets}

Due to the capabilities of the REST API, the tweets collected can be
older than the actual date of the event detected. Hence, 
some tweets can be very old and not relevant to the event itself. This
may lead to inaccuracies in predictions when using the early features.

\begin{figure}
  \centering
  \includegraphics[width=.7\textwidth]{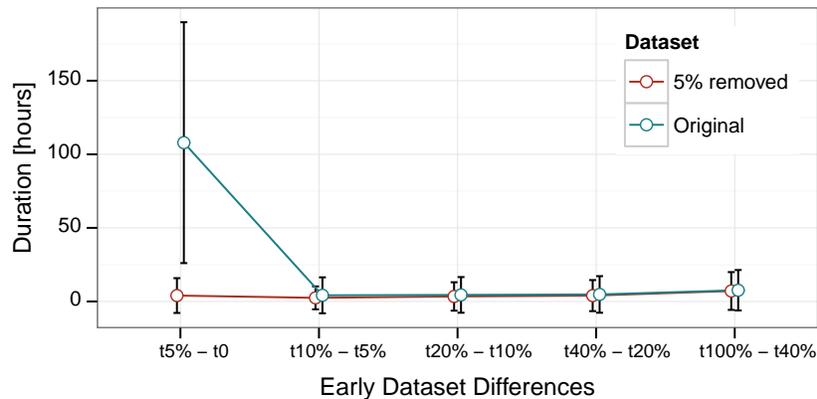}
  \caption[Duration differences of events.]{\textbf{Duration
      differences of events. The $x$-axis represents the categories of
      datasets: the first one (t5\%-t0) represents the difference of
      time between the timestamp of the oldest tweet and the newest
      tweet in the first 5\% of the tweets. The next one (t10\%-t5\%)
      corresponds to the difference between the newest tweet in the
      first 10\% and the newest tweet in the first 5\% of data, etc.
      After removing the first 5\% of data, the time differences are
      roughly the same across all datasets.
    }}\label{fig:duration-differences}

\end{figure}

This problem is illustrated in Figure~\ref{fig:duration-differences},
Note that the first 5\% of the tweets take an unusually
large portion of the duration of the entire event. This suggests that
we are collecting tweets which existed much before the event broke
out, and hence are possibly irrelevant. Once we discard the first 5\%
of tweets, we observe that each segment of the event (first 5\%, the
next 5\%, etc.) occupies roughly the same duration of the entire event.

\subsection{Validation of Data Collection}
\begin{figure}
  \begin{center}
    \includegraphics[width=0.7\textwidth]{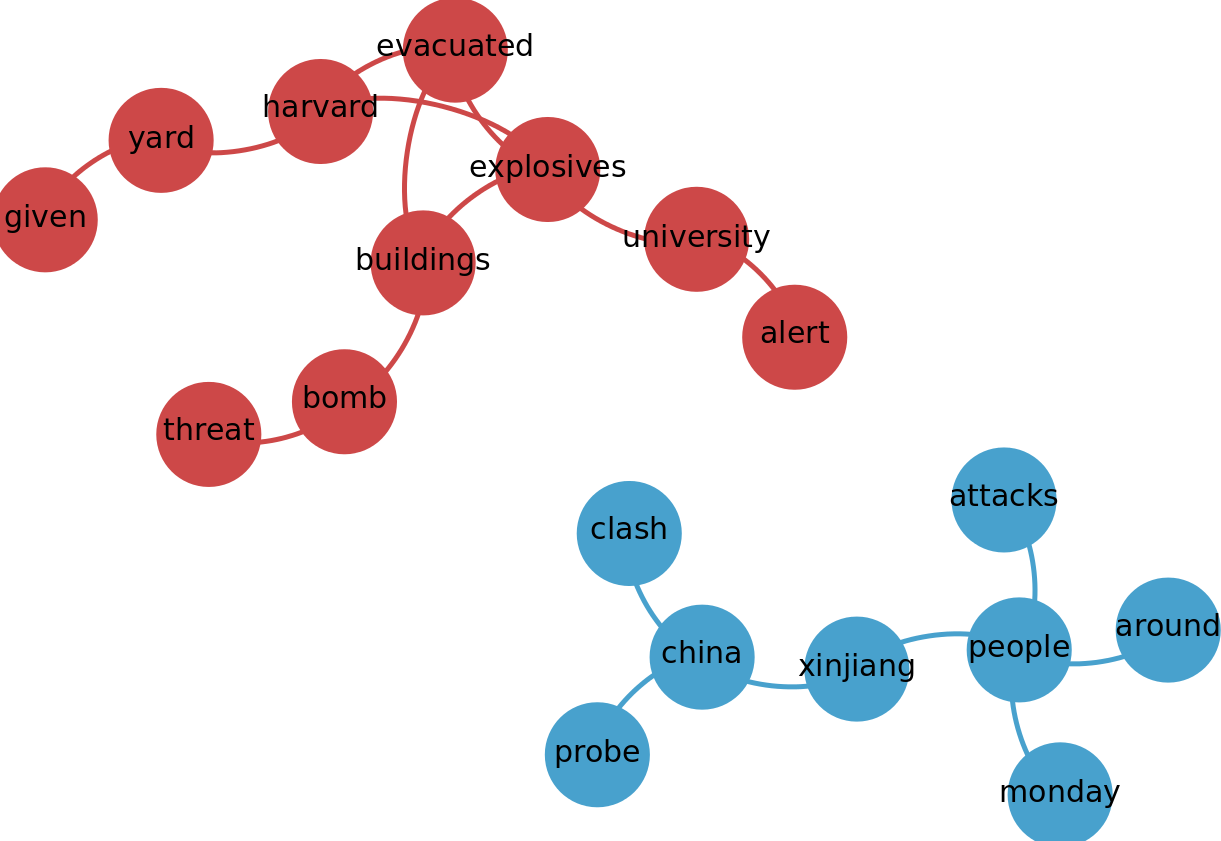}
    \caption{\textbf{This figure illustrates how we merge keyword
        pairs which represent the same event into larger components.
      }}
    \label{fig:connected_components}
  \end{center}
\end{figure}

\begin{figure}
  \centering
  \begin{subfigure}[b]{0.3\textwidth}
    \includegraphics[width=\textwidth]{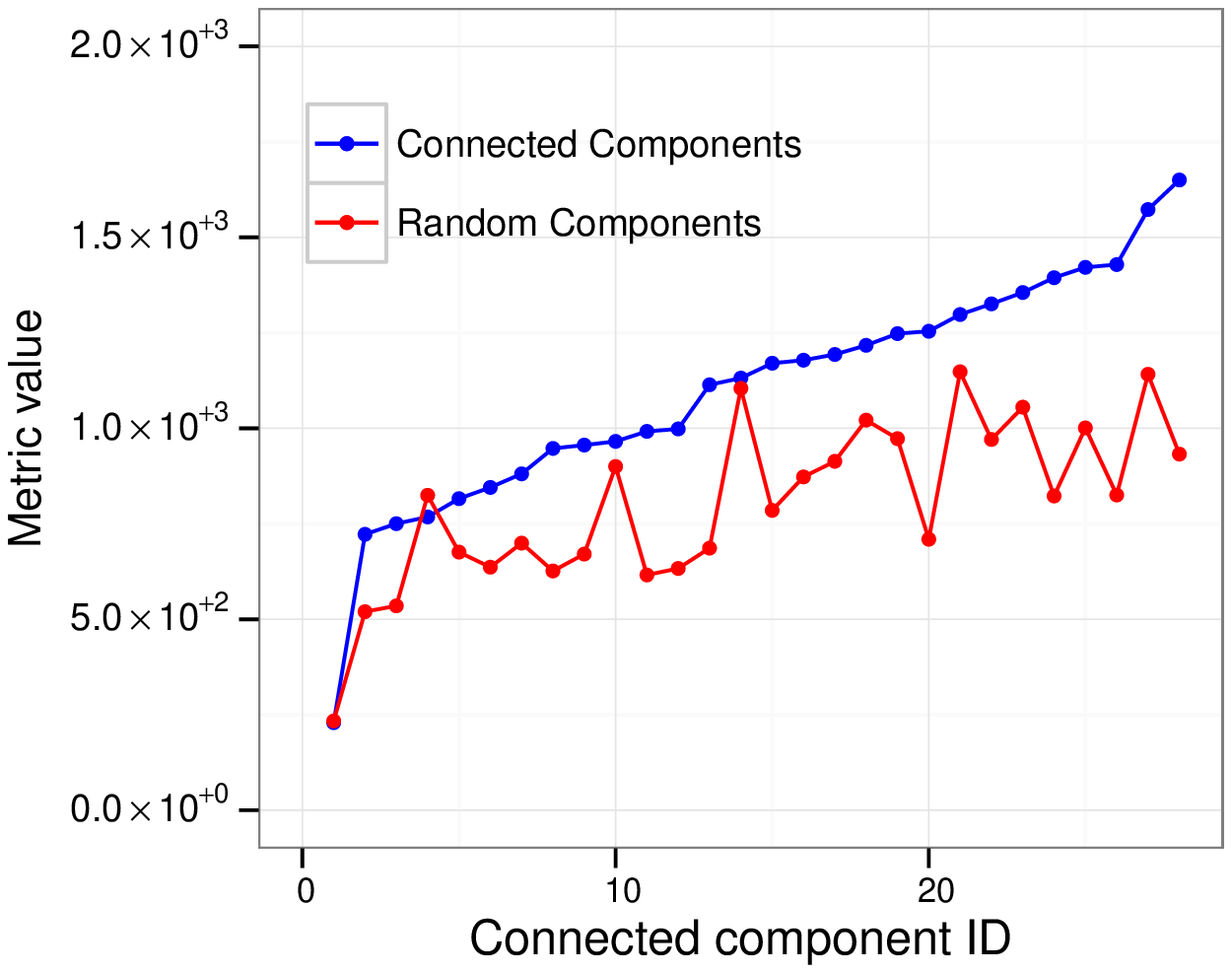}
    \caption{$I_1$} \label{fig:I1}
  \end{subfigure}
  \begin{subfigure}[b]{0.3\textwidth}
    \includegraphics[width=\textwidth]{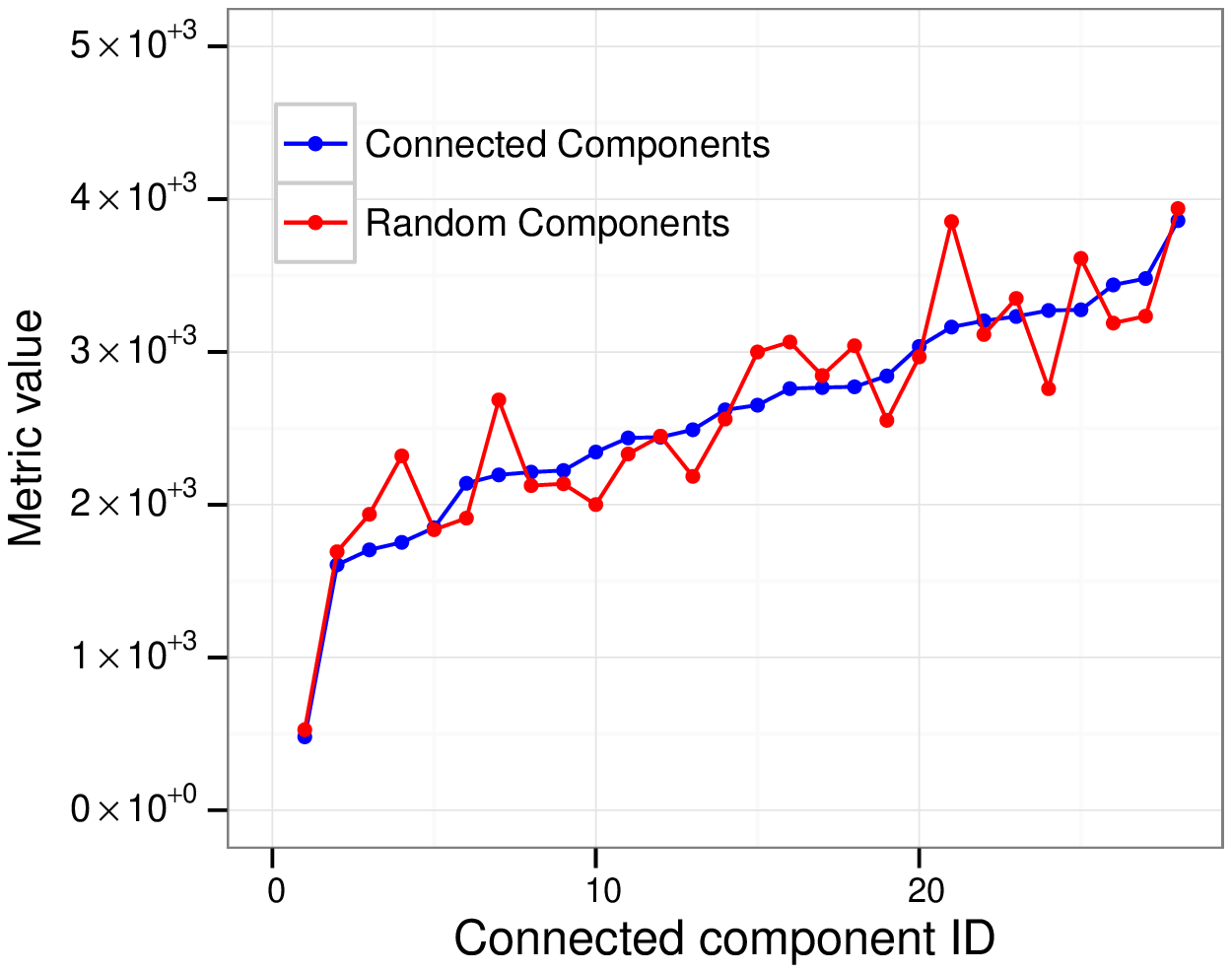}
    \caption{$I_2$} \label{fig:I2}
  \end{subfigure} ~ %add desired spacing
  \begin{subfigure}[b]{0.3\textwidth}
    \includegraphics[width=\textwidth]{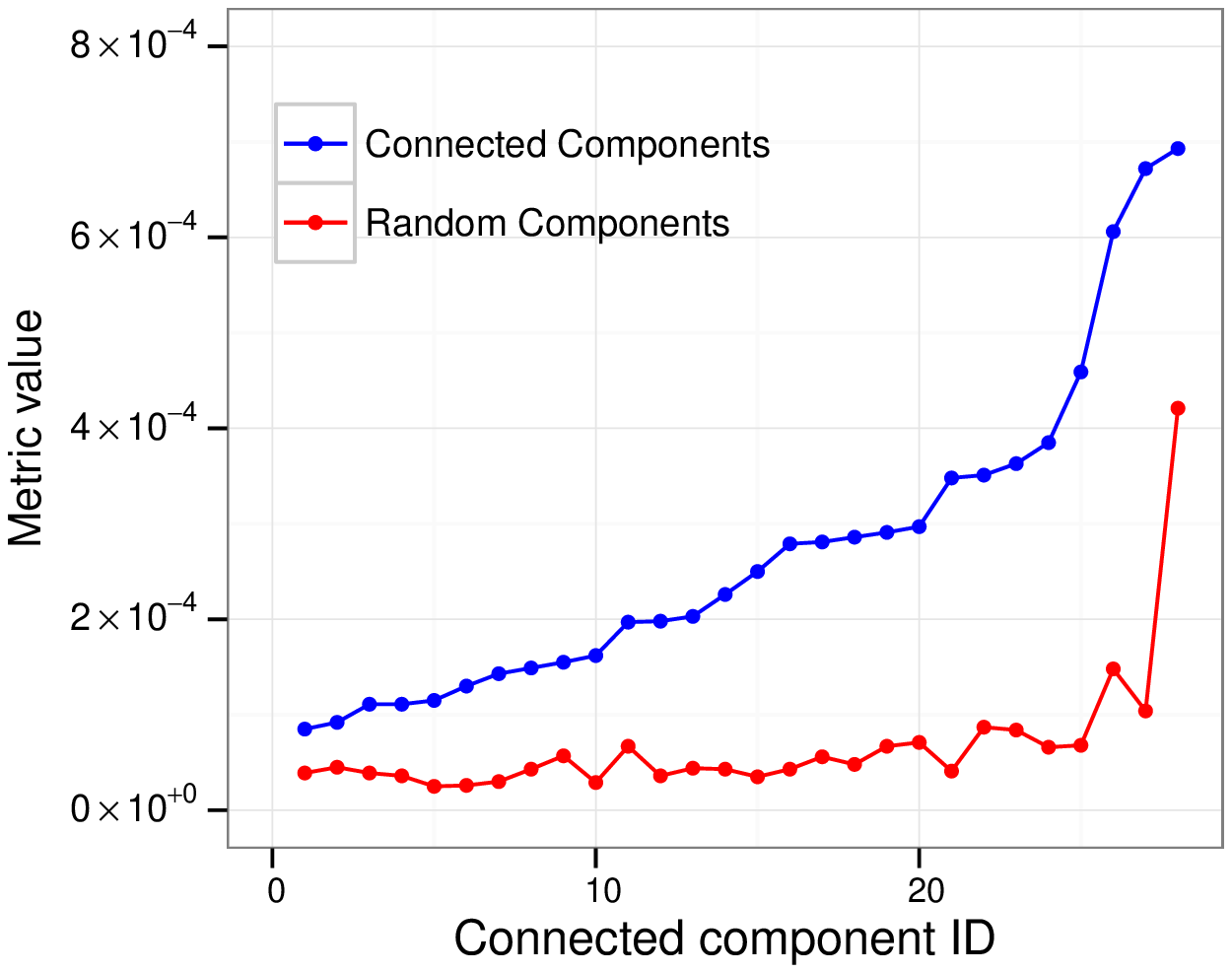}
    \caption{$H_1$} \label{fig:H1}
  \end{subfigure}

  \begin{subfigure}[b]{0.3\textwidth}
    \includegraphics[width=\textwidth]{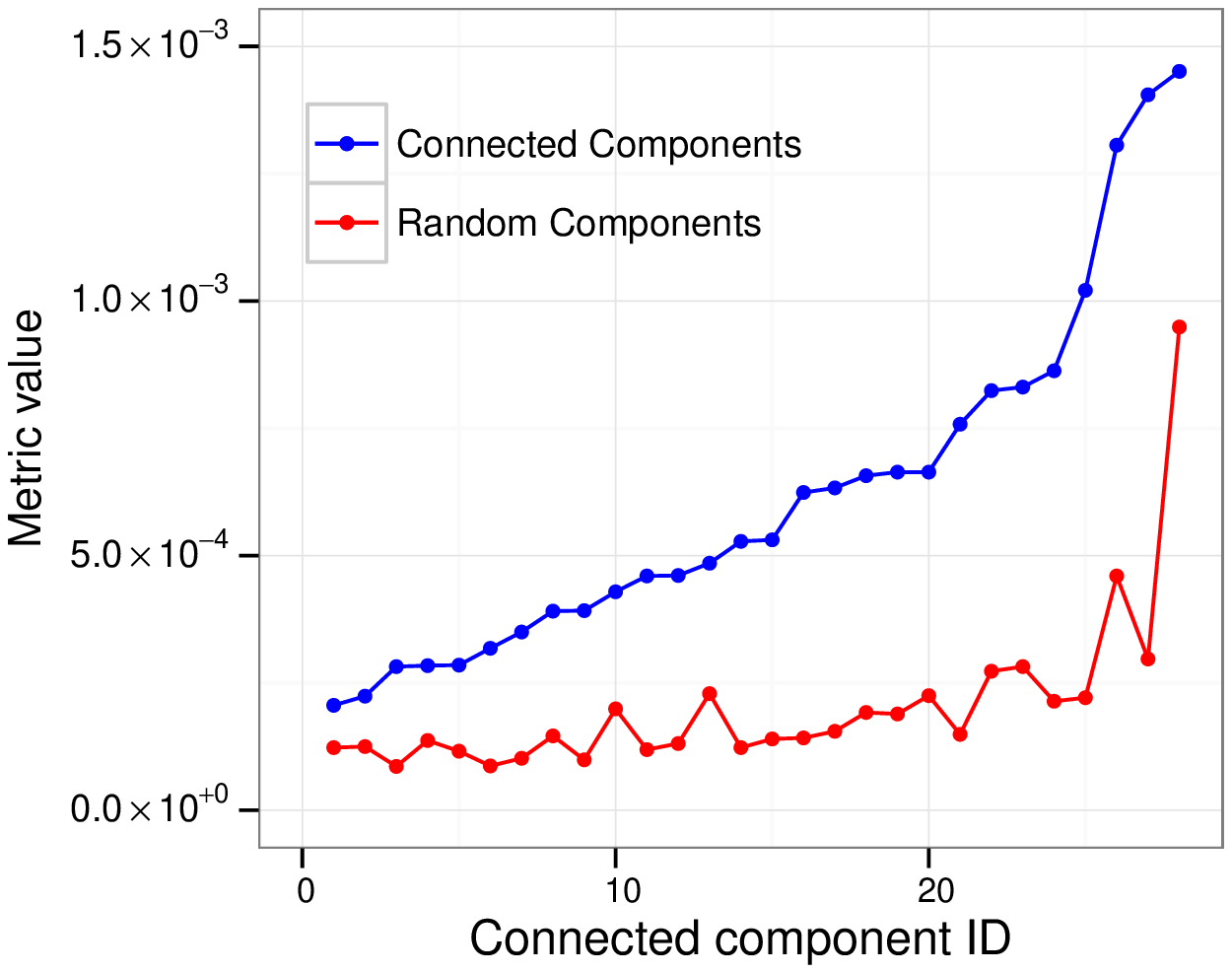}
    \caption{$H_2$} \label{fig:H2}
  \end{subfigure}
  \begin{subfigure}[b]{0.3\textwidth}
    \includegraphics[width=\textwidth]{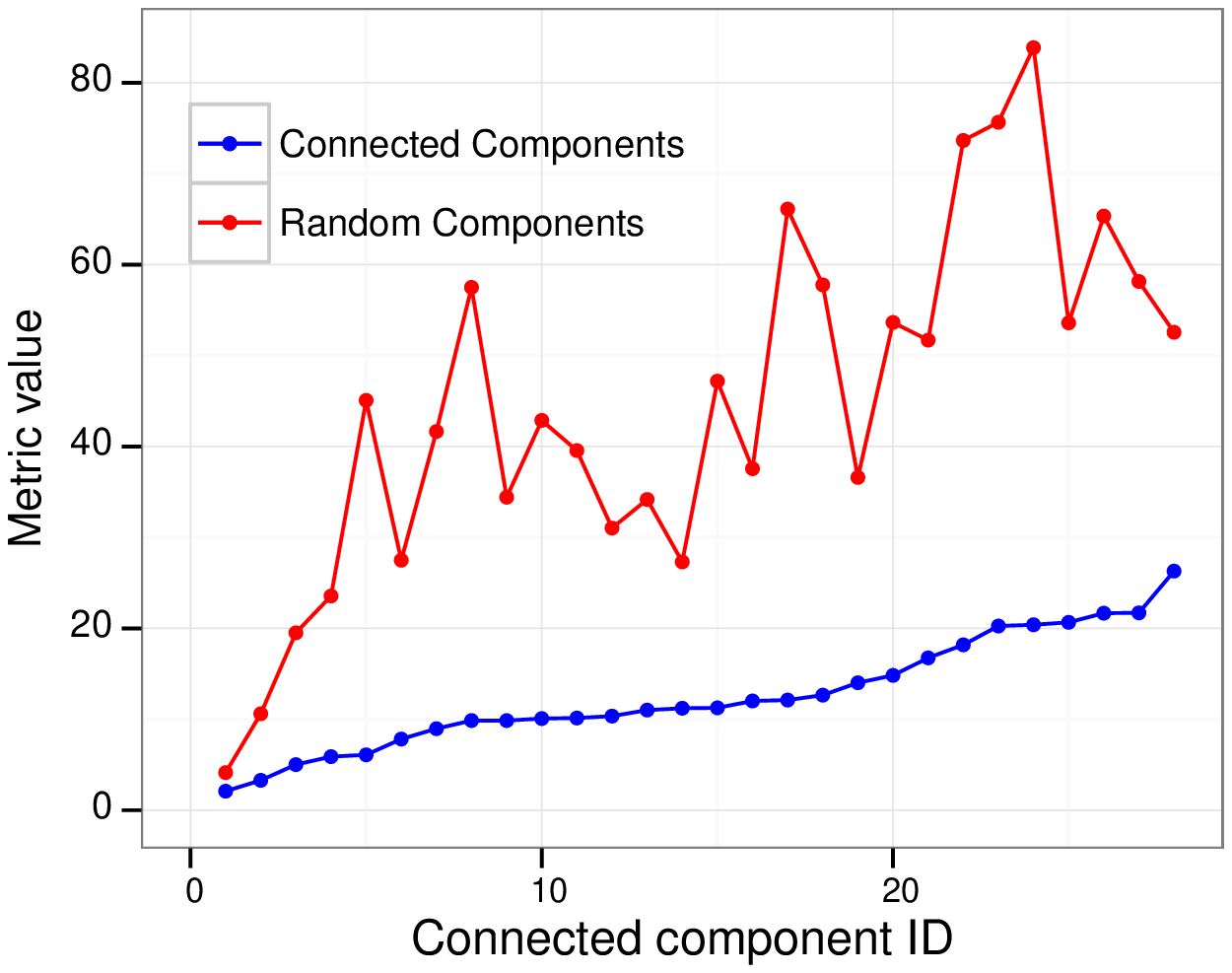}
    \caption{$G_1$} \label{fig:G1}
  \end{subfigure} ~ %add desired spacing
  \begin{subfigure}[b]{0.3\textwidth}
    \includegraphics[width=\textwidth]{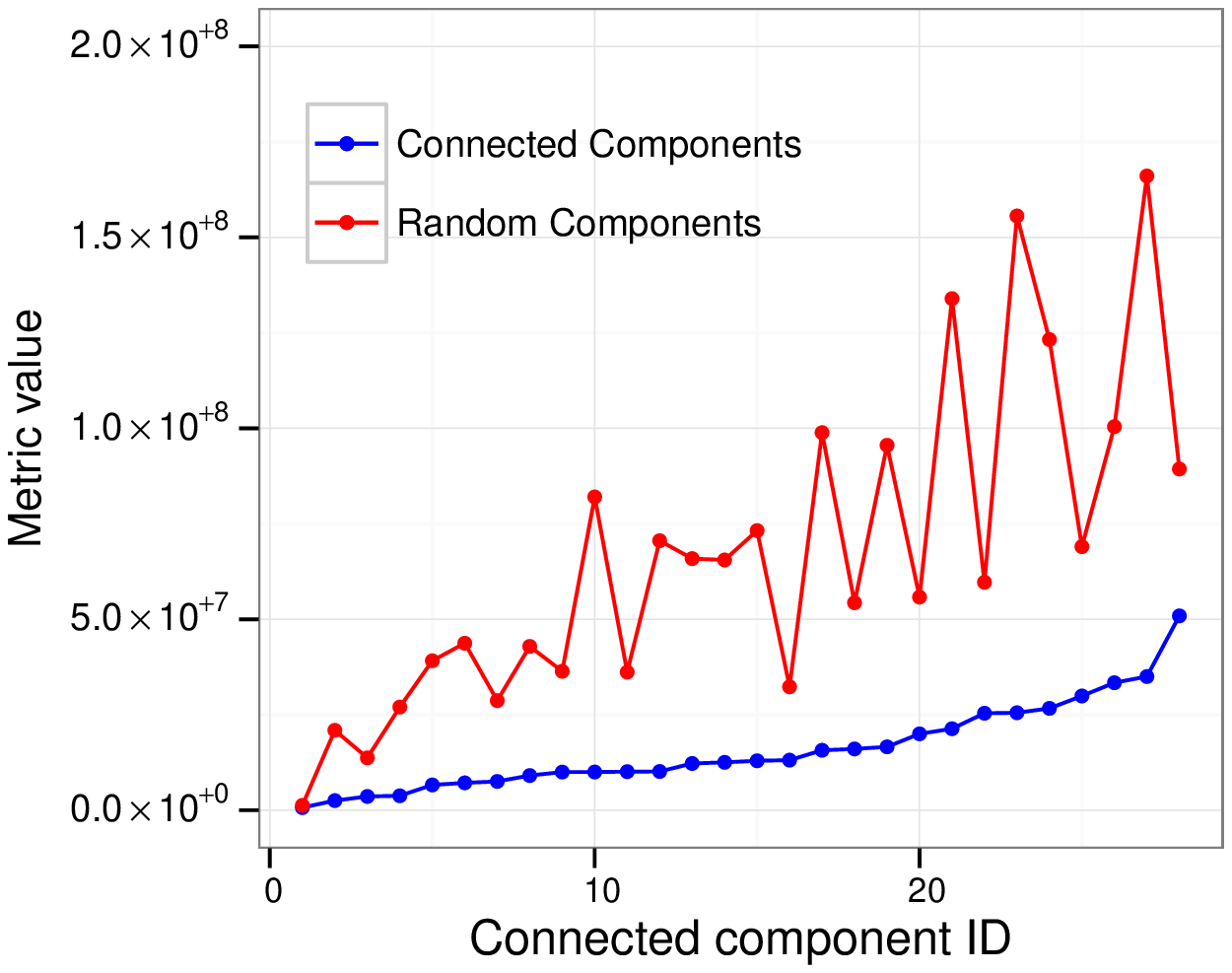}
    \caption{$G'_1$} \label{fig:G1P}
  \end{subfigure}

  \caption{\textbf{Each plot in this figure compares the
          quality of the cluster of tweets obtained from connected
          components and random components. The actual metric is shown
          in Table~\ref{table:clustering_metrics}. In $I_1, I_2, H_1,
          H_2$ higher value is better. In $G_1, G_1^{'}$, lower value
          is better. For visual clarity, the values obtained from
          connected components were sorted in ascending order, hence
          the blue line is monotonically increasing. The values
          obtained were rearranged in the same order as well.}}
    \label{fig:connected_components_validation}
\end{figure}

We performed experiments validating that merging keywords by forming
connected components indeed produced meaningful groups of keywords
representing an event. As a baseline, we used components obtained by
merging random keyword pairs together. We evaluated how well a cluster
is formed from the set of tweets obtained from connected components,
comparing the cluster to the set of tweets obtained from random
components. Connected components are expected to merge
keyword pairs that belong to the same event, and hence would make
better clusters when compared to merging random keyword pairs. The
results are displayed in
Figure~\ref{fig:connected_components_validation}. In this figure, each
plot depicts a different metric that evaluates the quality of a
cluster. These clustering metrics are summarized in
Table~\ref{table:clustering_metrics}. For better interpretation and
visual clarity, in each of the plots, we sorted the clustering metrics
obtained via connected components. We then rearranged the clustering
metrics for the baseline according to the sorting order obtained from
connected components. (This is the reason why the blue line is
monotonically increasing.) This experiment was performed on one month
of data (there are approximately 30 data points in each plot) between
August 2013 and September 2013. We took all the keyword pairs obtained
in a day and found the connected components as in
Figure~\ref{fig:connected_components}. For random components, we
merged the keyword pairs randomly. We took precautions to make sure
that the size of the connected components and random components per
day were comparable. That is, if we had connected components of sizes
6, 6, and 5 formed from keyword pairs on particular day, we made sure
that similarly sized random components were also formed from the
keyword pairs of the same day. Also, to make sure that tweets from any
one keyword pair do not dominate the tweet set, we sampled an equal
number of tweets from each keyword pair, and the \emph{same} sample of
tweets is used to calculate the clustering metrics in both the connected
components approach and the random components approach. The random
baseline has been averaged over 3 different rounds of experimentation.

%\begin{figure}
%  \begin{center}
%    \includegraphics[width=1\textwidth]{figures/Plots_from_data/connected_components_validation}
%    \caption{\textbf{\small{Each plot in this figure compares the
%          quality of the cluster of tweets obtained from connected
%          components and random components. The actual metric is shown
%          in Table~\ref{table:clustering_metrics}. In $I_1, I_2, H_1,
%          H_2$ higher value is better. In $G_1, G_1^{'}$, lower value
%          is better. For visual clarity, the values obtained from
%          connected components were sorted in ascending order, hence
%          the blue line is monotonically increasing. The values
%          obtained were rearranged in the same order as well.}}}
%    \label{fig:connected_components_validation}
%  \end{center}
%\end{figure}

\begin{table}
  \centering
  \begin{tabular}{ c  c  c }
    \toprule
    Name & Metric & Meaning \\
    \midrule
    $I_1$ & $\sum_{i=1}^k \frac{1}{n_i} \sum_{(u,v) \in S_i} \text{sim}(u,v)$ & Higher value is better\\
    \midrule
    $I_2$ & $\sum_{i=1}^k \sqrt{ \sum_{(u,v) \in S_i} (u,v)}$ & Higher value is better \\
    \midrule
    $E_1$ & $\sum_{i=1}^{k} n_i \frac{\sum_{v \in S_i, u \in S} \text{sim}(u,v)}{\sqrt{\sum_{(u,v) \in S_i} \text{sim}(u,v)}}$ & Lower value is better \\
    \midrule
    $G_1$ & $\sum_{i=1}^k \frac{\sum_{v \in S_i, u \in S}\text{sim}(u,v)}{\sum_{(v,u) \in S}\text{sim}(v,u)}$ & Lower value is better \\
    \midrule
    $G_1^{'}$ & $\sum_{i=1}^k n_i^2 \frac{\sum_{v \in S_i, u \in S}\text{sim}(v,u)}{\sum_{(u,v) \in S_i}\text{sim}(u,v)}$ & Lower value is better \\
    \midrule
    $H_1$ & $\frac{I_1}{E_1}$ & Higher value is better \\
    \midrule
    $H_2$ & $\frac{I_2}{E_1}$ & Higher value is better \\
    \bottomrule
  \end{tabular}
  \caption{\textbf{This table lists the clustering metrics used in Figure~\ref{fig:connected_components_validation}.}}
  \label{table:clustering_metrics}
\end{table}

\section{VQ Event Model}
%We propose a novel definition for the impact of a complex unit of
%information, in this case a news event, on the Web. This definition is
%motivated by the need to estimate the strength of the reaction in
%terms of activity, that an event produces in Online Social Networks.
%Impact should not be dependent on the size and duration of an event,
%allowing us to measure both {\em global} (with large network coverage)
%as well as {\em local} (with smaller network coverage) events.
%Following this motivation, we present a vector model for event impact
%based on the distribution of arrival interval rates between pairs of
%messages. Using this representation, we can group similar events
%together and identify clear separations between different types of
%events.

%This model is resistant to common issues that arise when estimating
%network impact based on the total number of messages of an event, or
%total number of shares, etc. For example, consider two events, one of
%1 hour duration, and another of 100 days duration. Assume that both
%events receive one hundred thousand shares in the first 20\% of their
%life-span, and very little subsequent attention. If we were to
%estimate their impact based on the number of shares of each event,
%they might seem the same. Even when normalizing by the total duration
%of each event, both events will appear equal. Nevertheless,
%intuitively the event with 1 hour duration should be considered of
%much higher-impact given the instancy of the reaction generated
%comparatively by the network.

%%% Details %%%%
We introduce a novel vectorial representation based on a vector quantization of the
interarrival time distribution, which we call ``VQ-event model".
The most representative interarrival times are learned from a large
training corpus.  Each of the learned interarrival times is called
a \emph{codeword}, and the entire set of the learned interarrival times, the \emph{codebook}.

We represent an event $e$, belonging to a collection of events
$\mathcal{E}$, as a tuple $(\mathcal{K}_e, \mathcal{M}_e)$, where
$\mathcal{K}_e$ is a set of \emph{keywords} and $\mathcal{M}_e$ is a
set of \emph{social media messages}. Both
the keywords and the messages are related to a real-world occurrence. 
As explained in Section The keywords are extracted in
order to succinctly describe the occurrence, and the messages are
posts from users about the event.

To learn the most representative interarrival times we perform
the following: for each $e \in
\mathcal{E}$ with messages $\mathcal{M}_e = \lbrack m_{1}^e, m_{2}^e,
\ldots m_{n}^e \rbrack$ and their corresponding time-stamps $\lbrack
t_{1}^e, t_{2}^e, \ldots t_{n}^e \rbrack$ where $t_{i} \leq t_{i+i}
\forall i \in [1,n]$, we compute all the interarrival times $d_{i}^e =
t_{i}^e-t_{i-1}^e$ (the value of $t_{0}$ is considered equal to
$t_{1}$ for initialization purposes). Then, the values of $d_{i}^e$
for all events in $\mathcal{E}$ are clustered to identify the {\em
  most representative} interarrival times.

Once the most representative interarrival times have been learned,
the vector quantizations for each event is produced as follows:
for each event, obtain all the interarrival times, and
quantize each of the interarrival times to the closest codeword
in the codebook.  This process is summarized in
Algorithm~\ref{alg:learn_representation}.
Line~\ref{alg:line:all_time_diff} collects all of the interarrival
times for all the events in $\mathcal{E}$ in
\textbf{f}. Line~\ref{alg:line:cluster} is a clustering algorithm
which takes \textbf{f} and the number of
clusters $k$ as inputs and returns the centroids of the clusters as
the output in \textbf{c}. The centroids can be thought of as the most
representative interarrival times for the event set $\mathcal{E}$. After that,
the interarrival times of each event $e$ is vector
quantized in terms of the centroids to obtain a $k$-dimensional real
valued representation of the event (Line~\ref{alg:line:vq}). 
In this representation, each entry is percentage of messages with that
particular codeword as the interarrival time.
\begin{algorithm}
  \caption{{\tt learn\_representation()}}
  \label{alg:learn_representation}
  \begin{algorithmic}[1]
    \REQUIRE Event set $\mathcal{E}$, and number of codewords $k$ in the codebook.\\
    \ENSURE A representation in $\mathbb{R}^{k}$ of each event $e = (\mathcal{K}_e, \mathcal{M}_e) \in \mathcal{E}$.\\
    \STATE $\textbf{f} \leftarrow \{d_{i}^e| m_{i}^e \in
    \mathcal{M}_e, e \in \mathcal{E} \}$
    \\ \label{alg:line:all_time_diff} \STATE \textbf{c} $\leftarrow$
    {\tt cluster(}$\textbf{f}, k${\tt)} \label{alg:line:cluster}
    \FOR{$e \in \mathcal{E}$} \STATE \textbf{e} $\leftarrow$ {\tt
      vq(}$d_{i}^e, \textbf{c}$ {\tt)}\\ \label{alg:line:vq}
    \ENDFOR
  \end{algorithmic}
\end{algorithm}

\section{High Activity Vs Low Activity Events}
\label{sec:diff}
Once the collection of events is converted into their VQ-event model representation, 
we can identify events that have produced similar levels of activity in the social
network. In other words, events are considered to have similar activity if the interarrival times
between their social media posts are similarly distributed, implying a very much alike
collective reaction from users to the events within a group. In order to identify groups of
similar events, we cluster the event models. We sort the resulting groups of events from
highest to lowest activity, according to the concentration of social media posts in the bins that
correspond to short interarrival times. We consider the events that fall in the top cluster to be
high-activity events as most of their interarrival times are concentrated in the smallest interval
of the VQ-event model.
Thus we end with four groups of events: high, medium-high, medium-low and low.
shows a heatmap of the interarrival relative frequency for each cluster.

Through this section, we analyze different features for each of the event categories and
compare them both qualitatively and quantitatively.  We peformed two-tailed $t$-tests
for a variety of features for events in the high-activity category, and compare it with
the average values for the remaining events. 

\subsection{Information Forwarding Characteristics}
\label{subsec:info_forwarding}
We found that the high-activity events possess more information
forwarding characteristics than other events. We present four features
which support this argument. The features, their description and their
values are listed in Table~\ref{tab:information_forwarding}.

The \texttt{retweet\_count} is generally higher for high-activity
events. This feature is the fraction of retweets present in the event,
log-normalized by the total amount of tweets in the event. A higher
value suggests that people have a greater tendency to spread the
occurrence of these events, and forward this information to their
followers. 

The \texttt{tweets\_retweeted} is lower for high-activity events than
for the rest. This feature is the number of tweets which have been
retweeted, log-normalized by the total number of tweets in the event.
This suggests that the high amount of retweets for the high-activity
events actually originates from fewer tweets.  This suggests that
fewer tweets become popular and are retweeted several times.

The \texttt{retweets\_most\_retweeted} is the total number of retweets
of the tweet that has been retweeted the most.  This number is much higher for high-activity
events than for low-activity events, suggesting that the most popular
tweet indeed becomes very popular when the event is of high-impact.

\begin{table}
  \centering
  {\scriptsize
    \begin{tabular}{llll}
      \toprule
      Feature Name &  \multicolumn{1}{l}{Description} & high-activity, others& Hypothesis, $p$-value\\
      \midrule
      \texttt{retweet\_count} & \pbox{20cm}{$\log($total retweet count \\in the event divided by total\\ tweets in the event$)$} & $2.205, 1.473$ & $1$, $p = 0$ \\
      \midrule
      \texttt{tweets\_retweeted} & \pbox{20cm}{$\log($number of tweets \\retweeted divided by\\ total tweets in the event$)$} & $-1.091, -0.964$ & $1$, $p = 2.7\times10^{-5}$ \\
      \midrule
      \texttt{retweets\_most\_retweeted} & \pbox{20cm}{number of tweets of the most \\retweeted tweet} & $284.491, 40.261$ & $1$, $p = 0$ \\
      \bottomrule
    \end{tabular}
  }
  \caption{\textbf{(Refer to Section~\ref{subsec:info_forwarding}.) This table lists all the features which characterize the information forwarding aspect of an event.  In general, high-activity events tend to have higher values for information forwarding features than other events.}}
  \label{tab:information_forwarding}
\end{table}

\subsection{Conversational Characteristics}
\label{subsec:conversational}
We found that high-activity events in general tend to generate more
conversation between users than the events in other categories. We observe this
behavior through several features. Refer to
Table~\ref{tab:conversational}.

\begin{table}
  \centering
  {\scriptsize
    \begin{tabular}{llll}
      \toprule
      Feature Name &  \multicolumn{1}{l}{Description} & high-activity, others & Hypothesis, $p$-value\\
      \midrule
      \texttt{replies} & \pbox{20cm}{$\log($total replies divided by total tweets$)$} & $-1.4016, -1.6474$ & $1$, $p = 10^{-4}$ \\
      \midrule
      \texttt{norm\_replies} & \pbox{20cm}{$\log($number of replies divided by\\ total number of unique users$)$} & $-1.5796, -1.9294$ & $1$, $p = 6.7\times10^{-4}$ \\
      \midrule
      \texttt{tweets\_replied} & \pbox{20cm}{$\log($number of tweets which generated\\ replies divided by total tweets$)$} & $-1.7784, -2.0668$ & $1$, $p = 0.001$ \\
      \midrule
      \texttt{uniq\_users\_replied} & \pbox{20cm}{$\log($unique users who have written\\ a reply divided by total tweets$)$} & $-1.7524, -2.0352$ & $1$, $p = 0.001$ \\
      \bottomrule
    \end{tabular}
  }
  \caption{\textbf{(Refer to Section~\ref{subsec:conversational}.)
      This table lists all the features which characterize the conversational aspect of high-activity and remaining events. Using
      these features, we argue in Section~\ref{subsec:conversational} that high-activity events tend to invoke more conversation amongst
      users than their counterparts.}}
  \label{tab:conversational}
\end{table}

The features \texttt{replies} and \texttt{norm\_replies} both count
the number of replies, but have been normalized slightly differently.
Both have a higher value for high-activity
events suggesting that high-activity events in general tend to spark
more conversation between the users. The \texttt{tweets\_replied}
feature counts the number of tweets which have generated replies (it
has been log-normalized by the total number of tweets in the event).
This is also higher for high-activity 
indicating that such events on average have more tweets which
invoke a reply from people. The \texttt{uniq\_users\_replied} feature
counts the number of unique users who have participated in an
conversation. Again, this number is found to be higher for high-activity
events than for others suggesting that more users tend to engage in a
conversation about these events. All these features collectively
suggest that high-activity events tend to have a \emph{conversational
  characteristic} associated with them.

\subsection{Topical Focus Characteristics}
\label{subsec:topical_focus}
We find that high-activity events have a lot more focus in terms of the
topical content than the remaining events. This possibly suggests that
when a news item is sensational, people seldom deviate from
the topic of the news to other things. 

We used four features listed in
Table~\ref{tab:topical_focus} to study the topic focus characteristics
of high-impact events.

\begin{table}
  \centering
  {\scriptsize
    \begin{tabular}{llll}
      \toprule
      Feature Name &  \multicolumn{1}{l}{Description} & high-activity, others & Hypothesis, $p$-value\\
      \midrule
      \texttt{uniq\_words} & \pbox{20cm}{$\log($total unique words \\divided by total tweets$)$} & $-0.1982, 0.1651$ & $1$, $p = 0$ \\
      \midrule
      \texttt{uniq\_chars} & \pbox{20cm}{$\log($total unique characters \\divided by total tweets$)$} & $2.0009, 2.0456$ & $1$, $p = 0$ \\
      \midrule
      \texttt{uniq\_hashtags} & \pbox{20cm}{$\log($number of unique hashtags\\ divided by total tweets$)$} & $-1.1126, 0.8761$ & $1$, $p = 0$ \\
      \midrule
      \texttt{uniq\_urls} & \pbox{20cm}{$\log($number of unique urls \\divided by total tweets$)$} & $-0.7194, -0.4951$ & $1$, $p = 0$ \\
      \bottomrule
    \end{tabular}
  }
  \caption{\textbf{(Refer to Section~\ref{subsec:topical_focus}.)
      This table summarizes all the features that were used to study the topical focus characteristics
      of high-activity events.}}
  \label{tab:topical_focus}
\end{table}

The number of unique words (\texttt{uniq\_words}) and characters
(\texttt{uniq\_chars}) for high-activity events is lower than the
remaining events suggesting that the information content for high-activity
events is more focused than for the remaining events (as they do not need
a diverse vocabulary). Hashtags on Twitter are a sequence of
characters that follow the \# symbol. Conventionally, their purpose is
to indicate the topic of the tweet. Again, this number
(\texttt{uniq\_hashtags}; log-normalized by the total number of
tweets) is lower for the high-activity events than for the remaining events.
The number of unique URLs (\texttt{uniq\_urls}; which can be
taken to interpret similar semantics as the hashtags) is also lower
for high-activity events than for the rest. 

\subsection{Early prediction of high-activity events}
\label{subsec:classification}
The results from sections \ref{subsec:info_forwarding}, \ref{subsec:conversational} and \ref{subsec:topical_focus} suggest that high-activity events differ
considerably from other events in terms of how they are received by the users
and in terms of the response they invoke from the network.

In the next phase, our goal is to supervised machine learning only the
early tweets of an event to predict whether an event will generate high-activity or not. 
A list of all the features used for classification is shown in Table \ref{tab:feats}.
The classification was carried using logistic regression provided by the Weka package.
The data was split approximately into $60-20-20$ of training, test and validation sets
and the results were averaged over 5 runs of experiments.

Table \ref{tab:classification_results} illustrates the prediction results
from the earliest 5\% of the tweets tweets, and from using all the tweets.
We the false positive rate using only the early tweets is almost
as good as the false positive rate using all the tweets.
The same observation holds for the metrics precision and ROC-area as well.
However, we observe an $18\%$ increase in the recall (0.455 to 0.540).
This suggests that some high-activity events
perhaps do not start displaying their unique characteristics
well enough in their early stages. 
\begin{table}
  \centering
  % {\scriptsize
  \begin{tabular}{lcc|cc}
    \toprule
    \multirow{2}{*}{ }& \multicolumn{2}{c}{Early 5\% Tweets} & \multicolumn{2}{c}{All Tweets} \\
    \midrule
    % \cmidrule{2-5} \cline{2-5}
    & high-activity & others & high-activity & others \\
    % \midrule
    high-impact & $194$ & $232$ & $230$ & $196$\\
    non-high-impact & $43$ & $4\,765$ & $47$ & $4\,761$ \\
    \bottomrule
  \end{tabular}
  % }
  \caption{\textbf{Confusion matrix while predicting the top 8\% of events
      as high-activity.  The predictions were made using the early 5\% of the tweets, and by using
      all the tweets from the event.}}
  \label{tab:confusion_matrix}
\end{table}
\begin{table}

  \centering
  {\small
    \begin{tabular}{lcccc|cccc}
      \toprule
      & \multicolumn{4}{c}{Early 5\% Tweets} & \multicolumn{4}{c}{All Tweets} \\
      \midrule
      & FP-Rate & Precision & Recall & ROC-area & FP-Rate & Precision & Recall & ROC-area \\
      % \midrule
      high-activity & 0.009 & 0.819 & 0.455 & 0.900 & 0.01 & 0.830 & 0.540 & 0.945 \\
      others & 0.545 & 0.954 & 0.991 & 0.900 &  0.460 & 0.960 & 0.990 & 0.945 \\
      \bottomrule
    \end{tabular}
  }
  \caption{\textbf{Classification results of detecting whether an event from the top 8\% is high-impact or not
      while predicting from features extracted from the earliest 5\% of the tweets and from all the tweets belonging to the event.}}
  \label{tab:classification_results}
\end{table}

{\footnotesize
  \begin{longtable}{l|l|l}

    \hline \textbf{Feature Name} & \textbf{Normalized By} &
    \textbf{Normalization
      Method} \\
    \hline
    \endfirsthead
    \multicolumn{3}{l}%
    {\tablename\ \thetable\ -- \textit{Continued from previous page}} \\
    \hline \textbf{Feature Name} & \textbf{Normalized By} &
    \textbf{Normalization
      Method} \\
    \hline
    \endhead
    \hline \multicolumn{3}{l}{\textit{Continued on next page}} \\
    \endfoot
    \hline
    \endlastfoot

    \texttt{component\_size}	&	 None	&	  \\
    \texttt{total\_seconds}	&	 \texttt{total\_tweets}	& $\log(x) - \log(y)$ \\
    \texttt{total\_tweets}	&	 None	&	  \\
    \texttt{total\_retweets}	&	 \texttt{total\_tweets}	&	 $\log(x) - \log(y)$ \\
    \texttt{total\_tweets\_retweeted}	&	 \texttt{total\_tweets}	&	 $\log(x) - \log(y)$ \\
    \texttt{retweets\_most\_retweeted}	&	 \texttt{total\_retweets}	&	 $\log(x) - \log(y)$ \\
    \texttt{total\_mentions}	&	 \texttt{total\_tweets}	&	 $\log(x) - \log(y)$ \\
    \texttt{total\_unique\_mentions} &
    \texttt{total\_mentions}	&	 $\log(x) - \log(y)$ \\
    \texttt{total\_tweets\_with\_mention}	&	 \texttt{total\_tweets}	&	 $\log(x) - \log(y)$ \\
    \texttt{total\_tweets\_with\_mostfrequent\_mention}	&	 \texttt{total\_tweets\_with\_mention}	&	 $\log(x) - \log(y)$ \\
    \texttt{total\_hashtags}	&	 \texttt{total\_tweets}	&	 $\log(x) - \log(y)$ \\
    \texttt{total\_unique\_hashtags}	&	 \texttt{total\_hashtags}	&	 $\log(x) - \log(y)$ \\
    \texttt{total\_tweets\_with\_hashtag}	&	 \texttt{total\_tweets}	&	 $\log(x) - \log(y)$ \\
    \texttt{total\_tweets\_with\_mostfrequent\_hashtag}	&	 \texttt{total\_tweets\_with\_hashtag}	&	 $\log(x) - \log(y)$ \\
    \texttt{total\_urls}	&	 \texttt{total\_tweets}	&	 $\log(x) - \log(y)$ \\
    \texttt{total\_unique\_urls}	&	 \texttt{total\_urls}	&	 $\log(x) - \log(y)$ \\
    \texttt{total\_tweets\_with\_url}	&	 \texttt{total\_tweets}	&	 $\log(x) - \log(y)$ \\
    \texttt{total\_tweets\_with\_mostfrequent\_url}	&	 \texttt{total\_tweets\_with\_url}	&	 $\log(x) - \log(y)$ \\
    \texttt{total\_unique\_verified\_users}	&	 \texttt{total\_verified\_users}	&	 $\log(x) - \log(y)$ \\
    \texttt{total\_verified\_users}	&	 \texttt{total\_tweets}	&	 $\log(x) - \log(y)$ \\
    \texttt{total\_unique\_users}	&	 \texttt{total\_tweets}	&	 $\log(x) - \log(y)$ \\
    \texttt{total\_replies} & \texttt{total\_unique\_users} &
    $\log(x) - \log(y)$ \\
    \texttt{total\_tweets\_first\_replied} & \texttt{total\_tweets} &
    $\log(x) - \log(y)$ \\
    \texttt{total\_unique\_users\_replied} &
    \texttt{total\_unique\_users} &
    $\log(x) - \log(y)$ \\
    \texttt{total\_tweets\_replied}	&	 \texttt{total\_tweets}	&	 $\log(x) - \log(y)$ \\
    \texttt{total\_words}	&	 \texttt{total\_tweets}	&	 $\log(x) - \log(y)$ \\
    \texttt{total\_unique\_words}	&	 \texttt{total\_words}	&	 $\log(x) - \log(y)$ \\
    \texttt{total\_characters}	&	 \texttt{total\_tweets}	&	 $\log(x) - \log(y)$ \\
    \texttt{total\_rt\_count}	&	 \texttt{total\_tweets}	&	 $\log(x) - \log(y)$ \\
    \texttt{total\_fav\_count}	&	 \texttt{total\_tweets}	&	 $\log(x) - \log(y)$ \\
    \texttt{total\_positive\_sentiment}	& \texttt{total\_tweets}	&	 $x / y$ \\
    \texttt{total\_negative\_sentiment}	&	 \texttt{total\_tweets}	&	 $x / y$ \\
    \hline

    \caption[List of features used for characterization]{\textbf{List
        of features used for characterization and classification. The
        ``Normalization Method'' column corresponds to the method used
        to normalize the value of the first column using the value of
        the second column. For example, the total number of retweets
        was normalized dividing it by the total number of tweets, and
        then taking the logarithm. Zero values were replaced by
        $10^{-8}$.}}
    \label{tab:feats}
  \end{longtable}

\end{document}